 \documentclass[sigconf]{acmart}

\AtBeginDocument{%
  }

\copyrightyear{2022}
\acmYear{2022}
\setcopyright{rightsretained}
\acmConference[ASE '22]{37th IEEE/ACM International Conference on Automated Software Engineering}{October 10--14, 2022}{Rochester, MI, USA}
\acmBooktitle{37th IEEE/ACM International Conference on Automated Software Engineering (ASE '22), October 10--14, 2022, Rochester, MI, USA}
\acmDOI{10.1145/3551349.3556916}
\acmISBN{978-1-4503-9475-8/22/10}

\usepackage[noend]{algpseudocode}
\usepackage{tikz}
\usetikzlibrary{automata,arrows,positioning}
\usepackage{multirow}
\usepackage{subfigure}
\usepackage{makecell}
\usepackage[algoruled,linesnumbered,noend]{algorithm2e}%
\usepackage{colortbl}
\usepackage{enumitem}
\usepackage{tcolorbox}

\usepackage{hhline}
\definecolor{mygray}{gray}{.85}

\usepackage{hyperref}

\newcommand{\mb}{\mathcal{N}}

\newcommand{\bs}[1]{{\mathbf{#1}}}

\newcommand{\myrom}[1]{\uppercase\expandafter{\romannumeral#1}}

\newcommand{\fu}[1]{\textcolor{red}{#1}}

\newcommand{\hide}[1]{ }

\newcommand{\tool}{{\sf QVIP}\xspace}

\begin{document}

\title[\tool: A Formal Verification Approach for QNNs]{\tool: An ILP-based  Formal Verification Approach \\ for Quantized Neural Networks}

\author{Yedi Zhang}
\email{zhangyd1@shanghaitech.edu.cn}
\orcid{0000-0003-1005-2114}
\affiliation{%
  \institution{ShanghaiTech University}
  \city{Shanghai}
  \country{China}
}
\affiliation{%
  \institution{Singapore Management University}
  \country{Singapore}
}

\author{Zhe Zhao}
\email{zhaozhe1@shanghaitech.edu.cn}

\author{Guangke Chen}
\email{chengk@shanghaitech.edu.cn}
\orcid{0000-0001-8277-3119}
\affiliation{%
  \institution{ShanghaiTech University}
  \city{Shanghai}
  \country{China}
}

\author{Fu Song}
\authornote{Corresponding author}
\email{songfu@shanghaitech.edu.cn}
\affiliation{%
  \institution{ShanghaiTech University}
  \city{Shanghai}
  \country{China}
}

\author{Min Zhang}
\email{zhangmin@sei.ecnu.edu.cn}
\orcid{0000-0003-1938-2902}
\affiliation{%
  \institution{East China Normal University}
  \city{Shanghai}
  \country{China}
}

\author{Taolue Chen}
\email{t.chen@bbk.ac.uk}
\orcid{0000-0002-5993-1665}
\affiliation{%
  \institution{Birkbeck, University of London}
  \city{London}
  \country{UK}
}

\author{Jun Sun}
\email{junsun@smu.edu.sg}
\orcid{0000-0002-3545-1392}
\affiliation{%
  \institution{Singapore Management University}
  \country{Singapore}
}

\renewcommand{\shortauthors}{Zhang et al.}

\begin{abstract}
Deep learning has become a promising programming paradigm in software development, owing to its surprising performance in solving many challenging tasks. Deep neural networks (DNNs) are increasingly being deployed in practice, but are limited on resource-constrained devices owing to their demand for computational power. Quantization has emerged as a promising technique to reduce the size of DNNs with comparable accuracy as their floating-point numbered counterparts. The resulting quantized neural networks (QNNs) can be implemented energy-efficiently. Similar to their floating-point numbered counterparts, quality assurance techniques for QNNs, such as testing and formal verification, are essential but are currently less explored. In this work, we propose a novel and efficient formal verification approach for QNNs. In particular,
we are the first to propose an encoding that reduces the verification problem of QNNs into the solving of integer linear constraints,
which can be solved using off-the-shelf solvers. Our encoding is both sound and complete. We demonstrate the application of our
approach on local robustness verification and maximum robustness radius computation. We implement our approach in a prototype tool \tool and conduct a thorough evaluation. Experimental results on QNNs with different quantization bits confirm the effectiveness and efficiency of our approach, e.g., two orders of magnitude faster and able to solve more verification tasks in the same time limit than the state-of-the-art methods.
\end{abstract}


\maketitle


\section{Introduction}\label{sec:intro}

Deep neural networks (DNNs) have gained widespread attention as a promising programming paradigm 
thanks to their surprising performance in solving various complicated tasks~\cite{KTSLSF14,HDYDMJSVNSK12,PSM14}.
DNNs are increasingly being deployed in software applications  
such as autonomous driving~\cite{Apollo} and medical diagnostics~\cite{CGGS12}, and have become a subject of intensive software engineering research.

Modern DNNs usually contain a great number of parameters which are typically stored as 32/64-bit floating-point numbers,
and require a massive amount of floating-point operations to compute the outputs at 
running time~\cite{TanL19}.
Hence, modern DNNs are both memory intensive and computationally intensive, making them difficult to deploy on energy- and computation-constrained devices~\cite{HanMD15}.
To address this issue, compression techniques have been proposed to `deflate' DNNs by removing redundant connections and quantizing parameters from 32/64-bit floating-points to lower bit-width fixed-points (e.g., 8-bits)~\cite{HanMD15,JacobKCZTHAK18}, which can vastly reduce the memory requirement and computational cost based on bit-wise and/or integer-only arithmetic computations.
For instance, Tesla's Full Self-Driving Chip (
previously Autopilot Hardware 3.0) is designed
for primarily  running 8-bit quantized DNNs~\cite{scaleQNN21,FSDChip}.
Quantization is also made available to ordinary programmers through popular deep learning frameworks. For instance,
TensorFlow Lite~\cite{TFLiteWebpage} supports, among others, 8-bit quantization on parameters.

Despite their great success, DNNs are known to be vulnerable to input perturbations due to the lack of
robustness~\cite{CW17b,PMJFCS16,goodfellow2014explaining,KGB17,SZSBEGF14,SongLCFL21,ChenCFDZSL21,DZBS21,CZSCFL22,chen2022towards,chen2021sec4sr}. This is concerning as such errors may lead to catastrophes when DNNs are deployed in safety-critical applications.
For example, a self-driving car may misclassify a stop sign as a speed-limit sign~\cite{EEF0RXPKS18}.
As a result, along with traditional verification and validation (V\&V) research in software engineering,
there is a large and growing body of work on developing V\&V techniques for DNNs, which has become a focus of software engineering researchers. For instance, testing techniques 
have been proposed to evaluate robustness of DNNs against input perturbations, e.g.,~\cite{CW17b,MJZSXLCSLLZW18,MaLLZG18,PCYJ17,SunWRHKK18,TianPJR18,XieMJXCLZLYS19,OdenaOAG19,XieMWLLL19},
cf.~\cite{ZhangHML22} for a survey. Such techniques are often effective in finding input samples to demonstrate lack of robustness, but they cannot prove the absence of such inputs. Many efforts have also been made to formally verify the robustness of
DNNs~\cite{PT10,KBDJK17,GMDTCV18,HKWW17,SGPV19,WPWYJ18,TranLMYNXJ19,ElboherGK20,YLLHWSXZ20,LiuSZW20,GuoWZZSW21,ZZCSCL22,LXSSXM22},
to cite a few. In general, we advocate the research on 
quality assurance for DNNs,
which are increasingly a component of modern software systems, to
which our current work contributes to.

Almost all the existing work is designated for  real or floating-point numbered DNNs only whereas verification of \emph{quantized} DNNs (QNNs) has not been thoroughly investigated yet,
although there is 
a gap of verification results between real-numbered DNNs and their quantized counterparts due to the fixed-point semantics of QNNs~\cite{GiacobbeHL20}.
Thus, there is a growing need for the research of quality assurance techniques for QNNs.
One possible approach to verify QNNs is to adopt differential verification~\cite{LahiriMSH13} which was initially proposed for 
verifying a new version of a program with respect to a
previous version. One could first verify a real or floating-point numbered DNN by applying existing techniques and then verify its quantized counterpart 
by applying differential verification techniques for 
DNNs~\cite{PaulsenWW20,PaulsenWWW20,MohammadinejadP21}.
However, it has two drawbacks. First, existing differential verification techniques for DNNs~\cite{PaulsenWW20,PaulsenWWW20,MohammadinejadP21} are incomplete,
and thus may produce false positives even if the original DNN is robust. 
Second, quantization introduces non-monotonicity on
the output of DNNs~\cite{GiacobbeHL20}, consequently, a robust DNN may become non-robust after quantization while
a non-robust DNN may become a robust QNN. Therefore, dedicated techniques are required for directly and rigorously verifying QNNs.

There do exist techniques for directly verifying QNNs which leverage Boolean Satisfiability (SAT) or Satisfiability Modulo Theory (SMT) solving or (reduced, ordered) binary decision diagrams (BDDs).
For 1-bit quantized DNNs, i.e., Binarized Neural Networks (BNNs),
Narodytska et al.~\cite{narodytska2018verifying} proposed to translate the BNN verification problem into the satisfiability problem of Boolean formulas where
SAT solving is harnessed. Using a similar encoding, Baluta, et al.~\cite{NPAQ19} proposed to quantitatively verify BNNs via approximate model counting. Following this direction,
Shih et al. \cite{ddlearning19A,ddlearning19B} proposed a BDD learning-based approach to quantitatively analyze BNNs,
and Zhang et al. \cite{BDD4BNN} introduced an efficient BDD encoding method by exploiting the internal structure of BNNs.
Recently, the SMT-based verification framework Marabou has also been extended to support 
BNNs~\cite{AWBK20}.
For quantization with multiple bits (e.g., fixed-point), 
methods also have been proposed~\cite{BaranowskiHLNR20,GiacobbeHL20,scaleQNN21}, which reduce the verification problem of QNNs to SMT solving accounting for the fixed-point semantics of QNNs. 
They are sound and complete, but have fairly limited scalability. 

In this work, we propose the first integer linear programming (ILP) based verification approach
for QNNs. 
To this end, we present a novel and exact encoding 
which precisely reduces
the verification problem for QNNs to an integer linear programming program.
More specifically, we propose to use piecewise constant functions to encode piecewise linear activation functions that
are computations of neurons in QNNs. The piecewise constant functions are further
encoded as integer linear constraints with the help of additional Boolean variables.
We also propose encodings to express desired input space and robustness properties using integer linear constraints.
The number of integer linear constraints produced by our encoding method is linear (at most 4 times) in
the number of neurons of QNNs. Based on our encoding, we develop
a theoretically complete and practically efficient verification framework for QNNs.
To further improve the scalability and efficiency, 
we leverage interval analysis to soundly approximate the neuron outputs,
which can effectively reduce the size of integer linear constraints and number of Boolean variables, and consequently, reduce verification cost.
We highlight two applications of our approach, i.e.,  robustness verification and computation of maximum robustness radii.



We implement our approach as an end-to-end tool, named \tool, with Gurobi~\cite{Gurobi} as the underlying ILP-solver.
We extensively evaluate \tool on various QNNs with different quantization bits using two popular datasets MNIST~\cite{MNIST} and Fashion-MNIST~\cite{Fashion-MNIST}, where the number of neurons varies from 858 to 894, and the number of bits for quantizing parameters
ranges from 4 to 10 bits with 8-bit input quantization.
For robustness verification, experimental results show that our approach is two orders of magnitude faster and is able to solve more verification tasks within the same time limit
than the state-of-the-art verifier for multiple-bit QNNs \cite{scaleQNN21}.
To the best of our knowledge, \tool is the first tool that can handle input spaces with an attack radius up to 30 for robust verification of QNNs.
Interestingly, we found that, although the accuracy of the QNNs stays similar under different quantization bits, the  robustness can be greatly
improved with more quantization bits using quantization-aware training~\cite{JacobKCZTHAK18}. We remark that Giacobbe et al.~\cite{GiacobbeHL20} showed that neither robustness nor non-robustness is monotonic with 
the number of quantization bits
using post-training quantization~\cite{nagel2021white}.
This suggests 
that robustness should also be considered, in addition to accuracy, during quantization-aware training 
which is able to produce more accurate QNNs than post-training quantization.
Furthermore, we show the effectiveness of \tool in computing maximum robustness radii based on binary search. Experimental results show that it
can be utilized to compare the overall robustness of QNNs with different quantization bits.

To summarize, our main contributions are as follows.

\vspace*{-1mm}

\begin{itemize}[leftmargin=*]
\item We propose the first ILP-based verification approach for QNNs featuring both precision and efficiency. 
\item We implement our approach as an end-to-end tool \tool, using the ILP-solver Gurobi for QNN robustness verification and maximum robustness radius computation.
\item We conduct an extensive evaluation of \tool, demonstrating the efficiency and effectiveness of \tool, e.g.,
significantly outperforming the state-of-the-art methods.
\end{itemize}
\vspace*{-0.5mm}

\noindent
{\bf Outline.} We define QNNs and problems in Section~\ref{sec:prel}.
In Section~\ref{sec:ilpEnc}, we propose the ILP-based verification approach. 
In Section~\ref{sec:maxR}, we present an algorithm for computing maximum robustness radii.
We report our experimental results in Section~\ref{sec:evaluation}. Section~\ref{sec:relatedwork} discusses related work. Finally, we conclude in Section~\ref{sec:conc}.

To foster further research, benchmarks and experimental data are
released on our website~\cite{website}. 

\section{Preliminaries}\label{sec:prel}

We denote by $\mathbb{R}$, $\mathbb{N}$, $\mathbb{Z}$ and $\mathbb{B}$ the set of real numbers, natural numbers, integers and Boolean domain $\{0,1\}$ respectively. Given a number $n\in \mathbb{N}$, let $[n]:= \{1,\cdots,n\}$, $\mathbb{R}^n$ and $\mathbb{Z}^n$ be the sets of the $n$-tuples of real numbers and integers, respectively. We use $\bs{W},\bs{W'},\ldots $ to denote matrices, $\bs{x},\bs{y}, \ldots$ to denote vectors, and $x,y, \dots$ to denote scalars.
We denote by $\bs{W}_{i,:}$ and $\bs{W}_{:,j}$ the $i$-th row and $j$-th  column of
the matrix $\bs{W}$. Similarly, we denote by $\bs{x}_j$ and $\bs{W}_{i,j}$
the $j$-th entry of $\bs{x}$ and $\bs{W}_{i,:}$ respectively.

\subsection{Deep Neural Networks}\label{subsec:qnnIntro}

\begin{figure}[t]
	\centering
	\subfigure[DNN $\mb_{e}$.]{\label{fig:dnnDemo}
		\begin{minipage}[b]{0.225\textwidth}
			\includegraphics[width=1.0\textwidth]{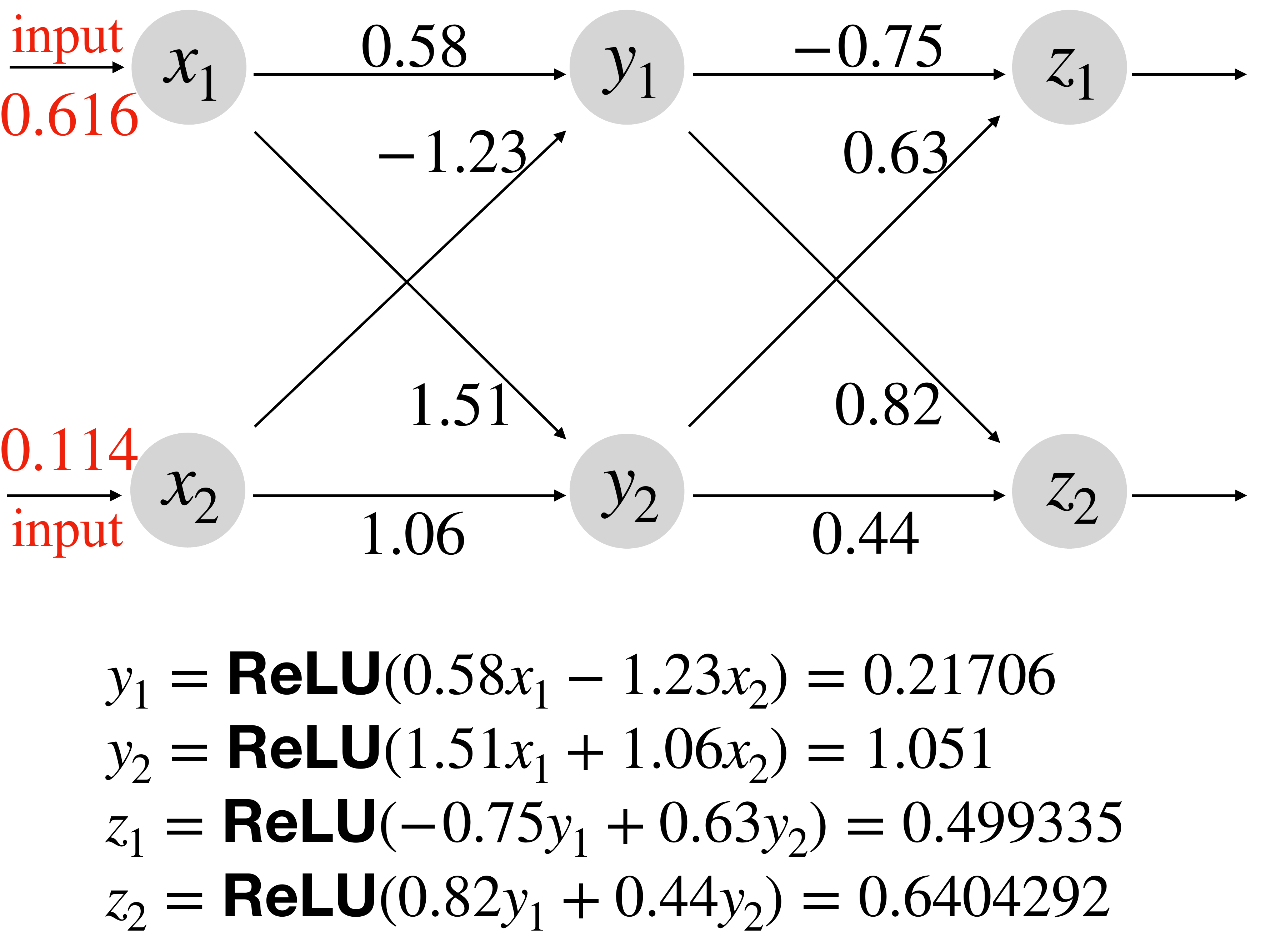}
		\end{minipage}	
	}
	\subfigure[QNN $\widehat{\mb}_{e}$.]{\label{fig:qnnDemo}
		\begin{minipage}[b]{0.225\textwidth}
			\includegraphics[width=1.0\textwidth]{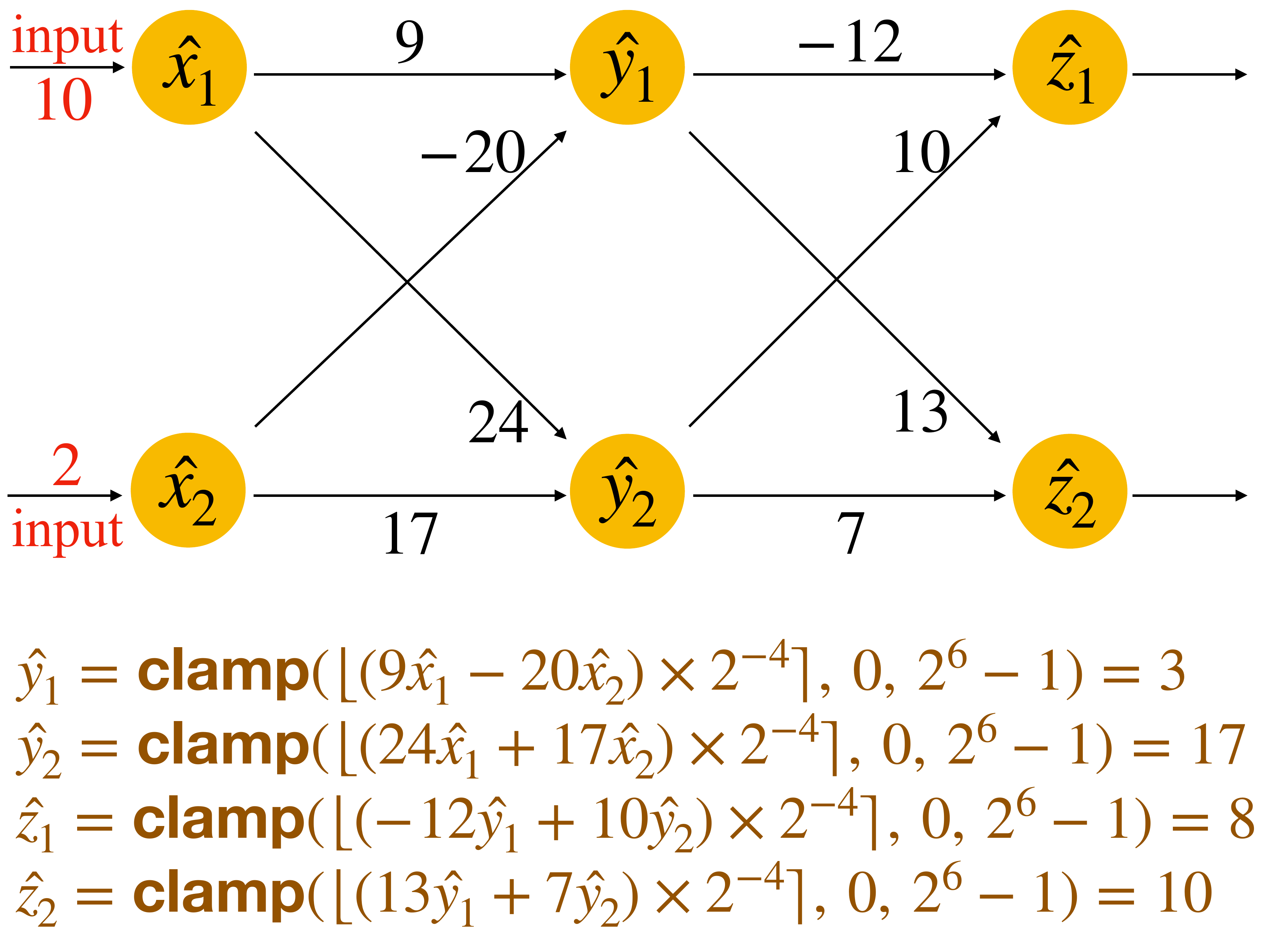}
		\end{minipage}	
	}
	\vspace*{-5mm}
	\caption{A 3-layer DNN ${\mb}_{e}$ and its quantized version $\widehat{\mb}_{e}$.}
	\vspace*{-3mm}
    \label{fig:nnVS}
\end{figure}

A deep neural network (DNN) $\mb$ is a graph structured in layers, where the first layer is the \emph{input layer}, the last layer is the \emph{output layer}, and the others are \emph{hidden layers}. All the nodes in these layers are called \emph{neurons} (in particular, neurons in hidden layers are called \emph{hidden neurons}). Each neuron in a non-input layer is associated with a \emph{bias} and could be connected by other neurons via weighted, directed edges. A DNN is \emph{feed-forward} if all the neurons only point to neurons in the next layer.
In this work, we focus on feed-forward DNNs. Given an input, a DNN computes an output by propagating it through the network layer by layer,
where the value of each neuron is computed by applying an \emph{activation function} to the weighted sum of output values from the preceding
layer or input.

A DNN $\mb$ with $d$ layers is a function $\mb: \mathbb{X} \rightarrow \mathbb{Y}$, where $\mathbb{X}\subseteq \mathbb{R}^n$
is the input domain and $\mathbb{Y}\subseteq \mathbb{R}^m$ is the output domain.
For any input $\bs{x}\in \mathbb{X}$, $\mb(\bs{x})= \bs{W}^d \bs{y}^{d-1} + \bs{b}^d$, where $\bs{y}^{d-1}$
is obtained by the following recursive definition:
\begin{center}
$\begin{aligned}
		\bs{y}^1 = \bs{x},  \qquad
		\bs{y}^i= \sigma(\bs{W}^i \bs{y}^{i-1} + \bs{b}^i)  \quad \mbox{for} \  i = 2, \ldots, d-1, \\
	\end{aligned}$
\end{center}
where $\bs{W}^i$ and $\bs{b}^i$ (for $2\leq i\leq d$)  are the weight matrix and bias vector of the $i$-th layer respectively,
and $\sigma$ is an activation function  (e.g., {\tt ReLU}, {\tt sigmoid}) applied to the vector entrywise.
In this work, we focus on the most commonly used one ${\tt ReLU}(x)=\max(x,0)$.

For classification tasks, the output \emph{class} of a given input $\bs{x}$, denoted by $\mb^c(\bs{x})$,   is the first index of $\mb(\bs{x})$ with the highest value.

\begin{example}\label{examp:dnn}
Figure~\ref{fig:dnnDemo} shows a running example DNN $\mb_{e}$ with 3 layers, where each layer has two neurons, the weights
are associated with the edges between neurons and all the biases are $0$.
For the input $\bs{x}=(0.616,0.114)$, the output of the first hidden layer is $\bs{y}=(0.21706,1.051)$
and the output of the DNN is $\bs{z}=(0.499335,0.6404292)$.
\end{example}

\subsection{Quantization of DNNs}
A quantized DNN (QNN) is structurally similar to its real-valued counterpart, except that the parameters, inputs of the QNN and outputs of each layer are quantized into integers, depending on
the given quantization configurations.

A \emph{quantization configuration} $\mathcal{C}$ is a tuple $\langle \tau, Q,F\rangle$, where $Q\in\mathbb{N}$
is the number of quantization bits for a value,
$F\in\mathbb{N}$ with $F\le Q$ is the number of quantization bits for the fractional part of the value,
and $\tau\in\{+,\pm\}$ indicates if the quantization is signed or unsigned.
The  configuration $\mathcal{C}$ defines the quantization grid limits $[\mathcal{C}^{\text{lb}},\mathcal{C}^{\text{ub}}]$,
where $\mathcal{C}^{\text{lb}}=-2^{Q-1}$ and $\mathcal{C}^{\text{ub}}=2^{Q-1}-1$ for $\tau=\pm$, $\mathcal{C}^{\text{lb}}=0$ and $\mathcal{C}^{\text{ub}}=2^Q-1$  for $\tau=+$.  
The quantized value $\hat{u}$ of a real value $u\in \mathbb{R}$ w.r.t.\ the quantization configuration $\mathcal{C}$ is defined as: 
\begin{center}
$\hat{u}=\text{clamp}\left(\big\lfloor 2^F\cdot u\big\rceil, \ \mathcal{C}^{\text{lb}}, \ \mathcal{C}^{\text{ub}}\right)$
\end{center}
where $\lfloor \cdot \rceil$ denotes the round-to-nearest integer operator and the clamping function $\text{clamp}(u',\alpha,\beta)$ with a lower bound $\alpha$ and an upper bound $\beta$ is defined as:
\begin{center}
$\text{clamp}(u',\alpha,\beta)=
\begin{cases}
  \alpha,  & \text{if} \ u'<\alpha;\\
  u',      &  \text{if} \ \alpha \le u'\le \beta;\\
  \beta,   &  \text{if} \ u'>\beta.
\end{cases}
$
\end{center}
%
Note that the quantized value $\hat{u}$ is an integer, which represents
the fixed-point value of $u$ in precision $\langle Q,F\rangle$. Thus, a QNN can be implemented in pure integer-arithmetic only.
For example, given a real value $u=1.2345$ and a quantization configuration $\mathcal{C}=\langle\pm, 4,2 \rangle$,
its quantized value $\hat{u}$ is $5$, representing its fixed-point value $1.25$.

Fix the quantization configurations
$\mathcal{C}_\bs{in}=\langle \tau_\bs{in}, Q_\bs{in},F_\bs{in}\rangle$,
$\mathcal{C}_\bs{w}=\langle \tau_\bs{w},  Q_\bs{w},F_\bs{w}\rangle$,
$\mathcal{C}_\bs{b}=\langle \tau_\bs{b}, Q_\bs{b},F_\bs{b}\rangle$,
and $\mathcal{C}_\bs{out}=\langle \tau_\bs{out},  Q_\bs{out},F_\bs{out}\rangle$
for inputs of the QNN, weights, biases
and outputs of each non-input layer, where $Q_\bs{in}-F_\bs{in}$, $Q_\bs{w}-F_\bs{w}$, $Q_\bs{b}-F_\bs{b}$ and $Q_\bs{out}-F_\bs{out}$
are large enough to represent the integer parts to avoid underflow and overflow during quantization.
Remark that each non-input layer can have its own quantization configurations $\mathcal{C}_\bs{w}$, $\mathcal{C}_\bs{b}$ and $\mathcal{C}_\bs{out}$ while we assume
all the non-input layers have the same quantization configurations for the sake of simplifying presentation.
 %

Given a weight matrix $\bs{W}$, a bias vector $\bs{b}$ and an input $\bs{x}$ of a DNN, their quantized versions $\widehat{\bs{W}}$, $\hat{\bs{b}}$ and $\hat{\bs{x}}$ w.r.t.\ $\mathcal{C}_\bs{w}$, $\mathcal{C}_\bs{b}$ and  $\mathcal{C}_\bs{in}$
are respectively defined as follows. For each $j,k$,
\begin{align*}
  \widehat{\bs{W}}_{j,k}&=\text{clamp}(\lfloor 2^{F_\bs{w}} \cdot \bs{W}_{j,k}\rceil, \ \mathcal{C}_\bs{w}^{\text{lb}}, \ \mathcal{C}_\bs{w}^{\text{ub}});\\
  \hat{\bs{b}}_{j}&=\text{clamp}(\lfloor 2^{F_\bs{b}} \cdot{\bs{b}}_{j} \rceil, \ \mathcal{C}_\bs{b}^{\text{lb}}, \ \mathcal{C}_\bs{b}^{\text{ub}});\\
  \hat{\bs{x}}_{j}&=\text{clamp}(\lfloor 2^{F_\bs{in}}\cdot \bs{x}_{j} \rceil, \ \mathcal{C}_\bs{in}^{\text{lb}}, \ \mathcal{C}_\bs{in}^{\text{ub}}).
\end{align*}

Given a $d$-layer DNN $\mb$, its \emph{quantized version} (i.e., QNN) w.r.t.
$\mathcal{C}_\bs{w}$, $\mathcal{C}_\bs{b}$, $\mathcal{C}_\bs{in}$ and  $\mathcal{C}_\bs{out}$ is defined as
the function $\widehat{\mb}:= \ell_{d}\circ \cdots \circ \ell_{1}$,
such that for every $i \in [d]$, $\ell_i:\mathbb{Z}^{n_i}\rightarrow\mathbb{Z}^{n_{i+1}}$ is a piecewise linear activation function, where
 \begin{itemize}[leftmargin=*]
  \item $n_1=n_2=n$, $n_{d+1}=m$;
  \item $\ell_{1}:\mathbb{Z}^{n_1}\rightarrow\mathbb{Z}^{n_{2}}$ is the identity function, i.e., $\hat{\bs{y}}^1=\ell_{1}(\hat{\bs{x}})=\hat{\bs{x}}$;

 %
\item $\ell_{i}$ for $2\leq i \leq d$ is the function such that
  for every ${\hat{\bs{y}}}^{i-1}\in \mathbb{Z}^{n_i}$ and  $j\in[n_{i+1}]$, the $j$-th entry of $\hat{\bs{y}}^i=\ell_i({\hat{\bs{y}}}^{i-1})$ is defined as
\begin{center}
$\hat{\bs{y}}_j^i=\text{clamp}\left(\Big\lfloor 2^{F_i}\sum\limits_{k=1}^{n_i}\widehat{\bs{W}}^i_{j,k}{\hat{\bs{y}}}_k^{i-1} + 2^{F_\bs{out}-F_\bs{b}} \hat{\bs{b}}_j^i \Big\rceil, \ \text{lb}, \ \mathcal{C}_\bs{out}^{\text{ub}}\right)$,
\end{center}
 where $F_i$ is $F_\bs{out}-F_\bs{in}-F_\bs{w}$ if $i=2$, and $-F_\bs{w}$ otherwise;
 $\text{lb}$ is $\mathcal{C}_\bs{out}^{\text{lb}}$
 if $i=d$, and  $0$ otherwise.
  \end{itemize}

Note that $2^{F_\bs{out}-F_\bs{in}-F_\bs{w}}$ and $2^{F_\bs{out}-F_\bs{b}}$ are used to align the precision between
the inputs and outputs of the first hidden layer, while $2^{-F_\bs{w}}$ and $2^{F_\bs{out}-F_\bs{b}}$
are used to align the precision between the inputs and outputs of the other hidden layers and output layer.
We notice that the ReLU activation function is avoided in the quantized hidden layers
by setting $\text{lb}$ to $0$ in the clamping functions for each hidden layer. Thus, for any input $\bs{x}\in \mathbb{R}^n$,
$\widehat{\mb}$ provides $\widehat{\mb}(\hat{\bs{x}})$.

\begin{example}\label{examp:qnn}
Consider the DNN $\mb_{e}$ given in Example~\ref{examp:dnn}.
The corresponding QNN $\widehat{\mb}_{e}$ w.r.t. the quantization configurations $\mathcal{C}_\bs{w}=\langle \pm,6,4\rangle$, $\mathcal{C}_\bs{in}=\mathcal{C}_\bs{out}=\langle +,6,4\rangle$
is shown in Figure~\ref{fig:qnnDemo}, where the quantized weighted are associated with the edges.
For the quantized input $\hat{\bs{x}}=(10,2)$ of $\bs{x}=(0.616,0.114)$ w.r.t. $\mathcal{C}_\bs{in}$, the output of the first hidden layer is $\hat{\bs{y}}=(3,17)$
and the output of $\widehat{\mb}_{e}$ is $\hat{\bs{z}}=(8,10)$.

%
\end{example}



\subsection{Robustness Problems}\label{subsec:qnnVeriDef}
In this work, we consider two robustness problems, i.e., (local) robustness and maximum robustness radius.

\noindent{\bf Robustness}.
Given a DNN $\mb: \mathbb{X} \rightarrow \mathbb{Y}$, an input $\bs{u}\in \mathbb{X}$, an attack radius $r\in\mathbb{R}$, and an $L_p$ norm for $p\in\{0,1,2,\infty\}$~\cite{CW17b},
 $\mb$ is \emph{robust} w.r.t. the input region $R_p(\bs{u},r)=\{ \bs{u}'\in\mathbb{R}^{n}\mid ||\bs{u}'-\bs{u}||_p\le r\}$, if
all the input samples from the region $R_p(\bs{u},r)$ have the same output (the same class for an classification task) as the input $\bs{u}$.
A sample $\bs{x}\in R_p(\bs{u},r)$ is called an \emph{adversarial example} of $\mb$ if $\mb(\bs{x})\neq \mb(\bs{u})$.

Similarly, we can define robustness of QNNs. Given a QNN $\widehat{\mb}: \mathbb{Z}^n \rightarrow \mathbb{Z}^m$ quantized from a DNN $\mb: \mathbb{X} \rightarrow \mathbb{Y}$, an input $\hat{\bs{u}}\in \mathbb{Z}^n$
quantized from an input $\bs{u}\in \mathbb{X}$, an attack radius $r\in\mathbb{N}$, and an $L_p$ norm,
the QNN $\widehat{\mb}$ is \emph{robust} w.r.t. the input region $\widehat{R}_p(\hat{\bs{u}},r)=\{ \hat{\bs{u}}'\in\mathbb{Z}^{n}\mid \|\hat{\bs{u}}'-\hat{\bs{u}}\|_p\le r\}$, if
all the input samples from the region $\widehat{R}_p(\hat{\bs{u}},r)$ have the same outputs (the same classes for classification tasks) as the input $\hat{\bs{u}}$.
Similarly, a sample $\hat{\bs{x}}\in \widehat{R}_p(\bs{u},r)$ is called an \emph{adversarial example} of $\widehat{\mb}$ if $\widehat{\mb}(\hat{\bs{x}})\neq \widehat{\mb}(\hat{\bs{u}})$.

It is known~\cite{GiacobbeHL20} that a DNN $\mb$ is not necessarily robust w.r.t. an input region $R_\infty(\hat{\bs{u}},r)$
even if its quantized version  $\widehat{\mb}$ is robust w.r.t. the input region $\widehat{R}_\infty(\hat{\bs{u}},r)$.
Similarly, a QNN $\widehat{\mb}$ is not necessarily robust w.r.t. an input region $\widehat{R}_\infty(\hat{\bs{u}},r)$
even if the original DNN $\mb$ is robust w.r.t. the input region $R_\infty(\hat{\bs{u}},r)$.
When QNNs are deployed in practice, it is
thus important to directly verify the robustness of QNNs instead of their original DNNs. Therefore, we focus on robustness verification of QNNs.

\smallskip
\noindent{\bf Maximum robustness radius (MRR)}.
Instead of verifying the robustness of a QNN w.r.t. an attack radius and an input $\hat{\bs{u}}\in\mathbb{Z}^{n}$, one may be interested in computing the maximum robustness radius $r\in\mathbb{N}$ such that the QNN is robust w.r.t. the attack radius $r$ and input $\hat{\bs{u}}$.
Given a QNN $\widehat{\mb}:\mathbb{Z}^{n}\rightarrow \mathbb{Z}^{m}$ and an input $\hat{\bs{u}}\in\mathbb{Z}^{n}$,
$r\in\mathbb{N}$ is the \emph{maximum robustness radius} (MRR) if the QNN $\widehat{\mb}$ is robust w.r.t. the input region $\widehat{R}_p(\hat{\bs{u}},r)$ but is not robust w.r.t.
 $\widehat{R}_p(\hat{\bs{u}},r+1)$.

MRR is an important metric for measuring the robustness of a QNN w.r.t. a set of selected inputs.
For instance, given the same classification problem with a set of selected inputs,
a QNN that has a larger average MRR is considered more robust than the one which has a smaller average MRR.

\section{Verification Approach} \label{sec:ilpEnc}
In this section, we propose a novel approach for directly verifying QNNs.
Our approach reduces the robustness verification problem of QNNs to an integer linear programming problem, which can be solved by off-the-shelf
solvers. We first show how to express piecewise constant functions as linear constraints, which is one of
the major building blocks of QNN encoding. We then present our QNN encoding as integer linear constraints by transforming piecewise linear activation functions of QNNs into piecewise constant functions.
Finally, we show how to express input regions and robustness properties as integer linear constraints.

%



\subsection{Encoding Piecewise Constant Functions}\label{sec:pwcf2ilp}
A \emph{piecewise constant function} $f:[a^\text{lb}, a^\text{ub})\rightarrow \mathbb{R}$ 
is defined as:
\begin{center}
$
   f(x) =
\begin{cases}
  t_1, &  \text{if } x\in [a_0, a_1);\\
  t_2, &  \text{if }  x \in [a_1, a_2);\\
  \vdots &  \vdots \\
  t_k, &  \text{if } x \in [a_{k-1}, a_k).
\end{cases}$
\end{center}
where $a_0=a^\text{lb}$, $a_k=a^\text{ub}$, $a_0,\cdots, a_{k}\in  \mathbb{R}$ with
$a_o<\cdots < a_k$. We note that $a^\text{lb}$ and $a^\text{ub}$ could be $-\infty$
and $+\infty$, respectively.

To express the above piecewise constant function $f$ in linear constraints, we introduce Boolean variables $v_1, \cdots,v_k$,
where for each $i\in[k]$, the variable $v_i$ is 1 (i.e., true) iff 
 $x\in [a_{i-1}, a_i)$. Note
that a Boolean variable $v$ can be seen as an integer variable with two additional constraints $v\geq 0$ and $v\leq 1$. Thus, Boolean variables
will be treated as integer variables in this work.

Let $\Psi_{f,x,y}$ be the following set of the linear constraints:
\begin{center}
$\Psi_{f,x,y}= \left\{
\begin{array}{l}
     v_1 +\cdots+v_k=1,  \     y=t_1 v_1+ \cdots + t_k v_k, \\
    x<a_1 v_1 +\cdots+a_{k-1} v_{k-1}+a_k' v_k, \\
    x \ge a_1 v_2+a_2 v_3+\cdots+ a_{k-1} v_{k}+a_0' v_1
\end{array}\right\}$
\end{center}
where $a_k'$ (resp. $a_0'$) is an extremely large (resp. small) integer number $\textbf{M}$ (resp. $-$\textbf{M})  if $a_k$
is $+\infty$ (resp. $-\infty$) otherwise $a_k$ (resp. $a_0$).
Clearly, $\Psi_{f,x,y}$ has 4 constraints and $k+2$ variables, where $x$ and $y$ denote the input and output of the function $f$.

Intuitively, $v_1 +\cdots+v_k=1$ ensures that there is exactly one Boolean variable $v_i$ whose value is $1$
and all the others are $0$, i.e., $x$ falls in \emph{one and only one} interval $[a_{i-1}, a_i)$ for some $1\leq i\leq k$;
$y=t_1 v_1+ \cdots + t_k v_k$ reformulates $f(x)=y$;
$x<a_1 v_1 +\cdots+a_k v_k$ ensures that $x <a_i$ if $v_i=1$,
while $x \ge a_1 v_2+a_2 v_3+\cdots+ a_{k-1} v_{k}+a_0 v_1$ ensures that
$x\geq a_{i-1}$ if $v_i=1$, thus $x\in [a_{i-1}, a_i)$ iff $v_i=1$ and $v_j=0$ for any $j\in[k]$ such that $j\neq i$.
We note that the variables $x$ and $y$ in $\Psi_{f,x,y}$ are real variables instead of integer variables.

\begin{proposition}\label{prop:pwcf2il}
For any $x\in [a^\text{lb},\ a^\text{ub})$ and $y\in\mathbb{R}$,
$\Psi_{f,x,y}$ holds  iff $f(x)=y$. \hfill $\qed$
\end{proposition}

\subsection{Encoding QNNs}\label{subsec:encQNN}
To encode a QNN as integer linear constraints, we
first transform  piecewise linear activation functions in the QNN into piecewise constant functions
which can be further expressed as sets of integer linear constraints by Proposition~\ref{prop:pwcf2il}.

Fix a QNN $\widehat{\mb}:= \ell_{d}\circ \cdots \circ \ell_{1}$ quantized from a DNN $\mb: \mathbb{X} \rightarrow \mathbb{Y}$
w.r.t. the quantization configurations $\mathcal{C}_\bs{in}$,
$\mathcal{C}_\bs{w}$,
$\mathcal{C}_\bs{b}$,
and $\mathcal{C}_\bs{out}$, where $\mathbb{X}\subseteq \mathbb{R}^n$,
$\mathbb{Y}\subseteq \mathbb{R}^m$, and for every $i \in [d]$, $\ell_i:\mathbb{Z}^{n_i}\rightarrow\mathbb{Z}^{n_{i+1}}$.

For the piecewise linear activation function $\ell_{1}$, it is the identify function, i.e., $\ell_{1}(\hat{\bs{x}})=\hat{\bs{x}}$,
thus, can be seen as a piecewise constant function.
We can directly build
a set $\Phi_1$ of integer linear constraints $\{\hat{\bs{y}}^1_j=\hat{\bs{x}}_j \mid j\in [n_1]\}.$
Clearly,  $\hat{\bs{y}}^1=\ell_{1}(\hat{\bs{x}})$ iff $\Phi_1$ holds.

It remains to handle the piecewise linear activation functions $\ell_i$ for $2\leq i\leq d$.
Recall that  for every ${\hat{\bs{y}}}^{i-1}\in \mathbb{Z}^{n_i}$ and  $j\in[n_{i+1}]$, the $j$-th entry of $\hat{\bs{y}}^i=\ell_i({\hat{\bs{y}}}^{i-1})$ is defined as
\begin{center}
$\hat{\bs{y}}_j^i=\text{clamp}\left(\Big\lfloor 2^{F_i}\sum\limits_{k=1}^{n_i}\widehat{\bs{W}}^i_{j,k}{\hat{\bs{y}}}_k^{i-1} + 2^{F_\bs{out}-F_\bs{b}} \hat{\bs{b}}_j^i \Big\rceil, \ \text{lb}, \ \mathcal{C}_\bs{out}^{\text{ub}}\right)$,
\end{center}
 where $F_i$ is $F_\bs{out}-F_\bs{in}-F_\bs{w}$ if $i=2$, otherwise $-F_\bs{w}$; $\text{lb}$ is $\mathcal{C}_\bs{out}^{\text{lb}}$
 if $i=d$, otherwise $0$.

For every $i:2\leq i\leq d$ and $j\in[n_{i+1}]$,
we define the following piecewise constant function $f_{i,j}:[-\infty,+\infty)\rightarrow \mathbb{Z}$:
\begin{center}
$
    f_{i,j}(\bs{z}^i_j) =
\begin{cases}
  t_1, & \text{if}\ \bs{z}^i_j\in [ -\infty ,  t_1+0.5) \\
  t_2, & \text{if}\ \bs{z}^i_j\in [ t_2-0.5 ,  t_2+0.5)\\
  \vdots &\vdots \\
  t_k, &  \text{if} \ \bs{z}^i_j\in [ t_k-0.5 , +\infty)
\end{cases}$
\end{center}
where $k=\mathcal{C}_\bs{out}^{\text{ub}}-\text{lb}+1$, $t_\xi =\text{lb}+\xi-1$ for $\xi\in [k]$.
 We note that +0.5 and -0.5 are added to ensure that $\lfloor \bs{z}^i_j \rceil =t_j$ iff $\bs{z}^i_j\in [ t_j-0.5 ,  t_j+0.5)$.
%
 Clearly,  we have:
 $f_{i,j}(\bs{z}^i_j) = \mbox{clamp}\left(\lfloor \bs{z}^i_j\rceil, \ \text{lb}, \ \mathcal{C}_\bs{out}^{\text{ub}}\right)$.
We denote by
\begin{itemize}[leftmargin=*]
  \item $\Psi_{f_{i,j}, \bs{z}^i_j,\hat{\bs{y}}_j^i}$ the set of 4 linear constraints
encoding the piecewise constant function $f_{i,j}$, involving  $\mathcal{C}_\bs{out}^{\text{ub}}-\text{lb}+3$ variables, where $\bs{z}^i_j$ and $\hat{\bs{y}}_j^i$ denote
the input and output of the function $f_{i,j}$;
  \item $\Phi_i$  the set $\bigcup_{j=1}^{n_{i+1}} \Psi_{f_{i,j}, \bs{z}^i_j,\hat{\bs{y}}_j^i}'$ that has $4 n_{i+1}$ constraints and $(\mathcal{C}_\bs{out}^{\text{ub}}-\text{lb}+2)n_{i+1}+n_i$ integer and Boolean variables, where
  $\Psi_{f_{i,j}, \bs{z}^i_j,\hat{\bs{y}}_j^i}'$ is obtained from $\Psi_{f_{i,j}, \bs{z}^i_j,\hat{\bs{y}}_j^i}$ by replacing each occurrence of $\bs{z}^i_j$ by 
  $2^{F_i}\sum\limits_{k=1}^{n_i}\widehat{\bs{W}}^i_{j,k}{\hat{\bs{y}}}_k^{i-1} + 2^{F_\bs{out}-F_\bs{b}} \hat{\bs{b}}^i_j$ for all $2\leq i\leq d$ and $j\in[n_{i+1}]$, so that
  $\Phi_i$ becomes the set of integer linear constraints;

  \item  $\Phi_{\widehat{\mb}}$  the set $\bigcup_{i=1}^d \Phi_i$ that has $\sum_{j=2}^{d} 4 n_{i+1}$ constraints and $n_1+(\mathcal{C}_\bs{out}^{\text{ub}}-\text{lb}+2)\sum_{i=2}^d n_{i+1}$ variables. (Note that
  $\Phi_1=\{\hat{\bs{y}}^1_j=\hat{\bs{x}}_j \mid j\in [n_1]\}$ is eliminated by replacing $\hat{\bs{y}}^1_j$ with $\hat{\bs{x}}_j$.)
\end{itemize}

%

From $f_{i,j}(\bs{z}^i_j) = \mbox{clamp}\left(\lfloor \bs{z}^i_j\rceil,  \text{lb},  \mathcal{C}_\bs{out}^{\text{ub}}\right)$ and Proposition~\ref{prop:pwcf2il}, we get that
  $\Psi_{f_{i,j},\bs{z}^i_j,\hat{\bs{y}}^i_j}$ holds iff $\hat{\bs{y}}^i_j=f_{i,j}(\bs{z}^i_j)$, $\Psi_{f_{i,j},\bs{z}^i_j,\hat{\bs{y}}^i_j}'$ holds iff $\hat{\bs{y}}^{i}_j=\ell_i(\hat{\bs{y}}^{i-1})$,
  and $\Phi_i$ holds iff $\hat{\bs{y}}^{i}=\ell_i(\hat{\bs{y}}^{i-1})$. Therefore, we have:

 \begin{proposition}\label{prop:qnn2il}
  $\Phi_{\widehat{\mb}}$ holds iff $\hat{\bs{y}}^{d}=\widehat{\mb}(\hat{\bs{x}})$, where the number of constraints  (resp. variables) of $\Phi_{\widehat{\mb}}$
  is at most $4$ (resp. $\mathcal{C}_\bs{out}^{\text{ub}}-\mathcal{C}_\bs{out}^{\text{lb}}+2$) times of the number of neurons of $\widehat{\mb}$.  \hfill $\qed$
 \end{proposition}

\begin{figure}[t]
	\centering
	\includegraphics[width=.4\textwidth]{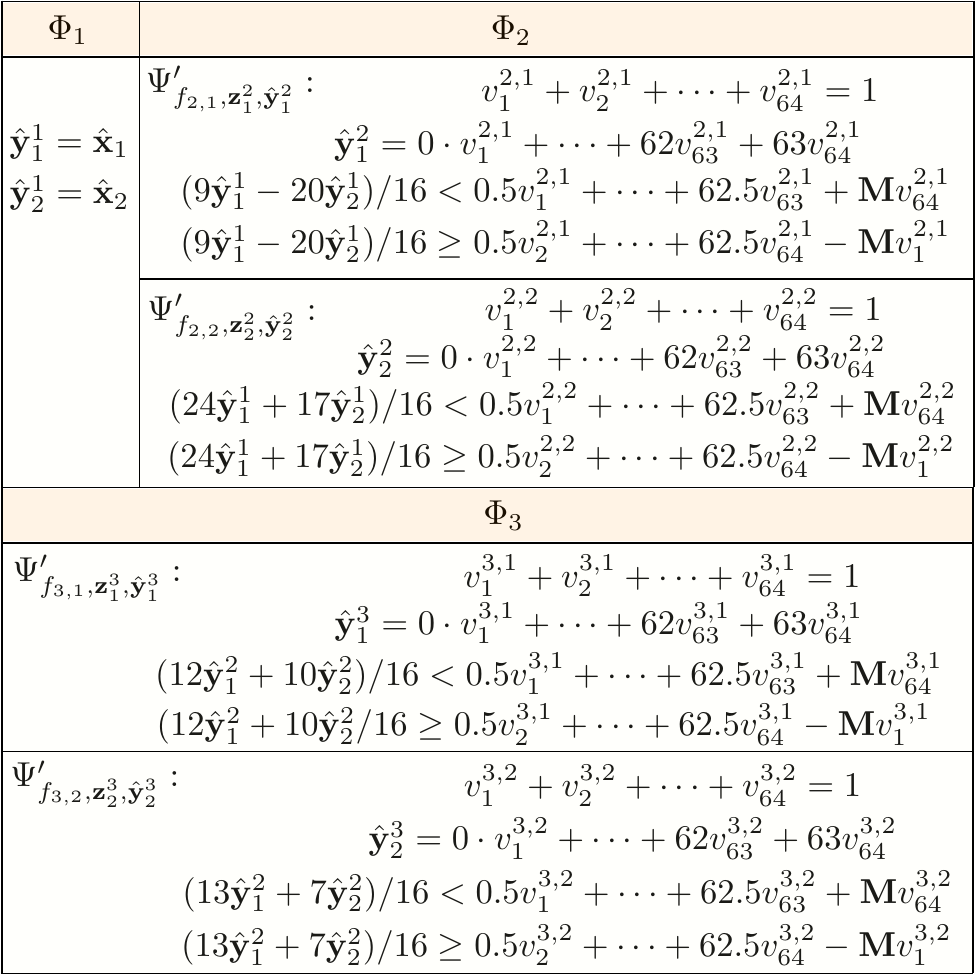}
  \vspace*{-2mm}
	\caption{An illustrating encoding of the QNN $\widehat{\mb}_{e}$.}\label{fig:qnnILPdemo}
  \vspace*{-2mm}
\end{figure}

\begin{example}\label{examp:qnn_ILP}
    Consider the QNN $\widehat{\mb}_{e}$ given in Example~\ref{examp:qnn}, quantized from $\mb_{e}$
    w.r.t. $\mathcal{C}_\bs{w}=\langle \pm,6,4\rangle$ and $\mathcal{C}_\bs{in}=\mathcal{C}_\bs{out}=\langle +,6,4\rangle$.
We have: $\text{lb}=0$ and $\mathcal{C}_\bs{out}^\text{ub}=2^6-1$ for both $\ell_2$ and $\ell_3$.
    The encoding of $\widehat{\mb}_{e}$ is shown in Figure~\ref{fig:qnnILPdemo}, where
 Boolean variables $\{v_1^{i,j},\ldots,v_{64}^{i,j}\mid 2\leq i\leq 3, 1\leq j\leq 2\}$ are introduced for encoding the piecewise linear activation functions
 $\ell_2$ and $\ell_3$.
\end{example}

\subsection{Encoding Input Regions}\label{subsec:encIR}
Recall that an input region $\widehat{R}_p(\hat{\bs{u}},r)=\{ \hat{\bs{u}}'\in\mathbb{Z}^{n}\mid ||\hat{\bs{u}}'-\hat{\bs{u}}||_p\le r\}$ is formed by an input $\hat{\bs{u}} \in\mathbb{Z}^{n}$, an attack radius $r\in \mathbb{N}$
 and an $L_p$ norm for $p\in\{0,1,2,\infty\}$.
We encode $\widehat{R}_p(\hat{\bs{u}},r)$ as a set $\Phi_{\hat{\bs{u}},r}^p$ of constraints:
\begin{itemize}[leftmargin=*]
%
  \item $\Phi_{\hat{\bs{u}},r}^0=\bigcup_{i \in [n]}\Psi_{g_i,\hat{\bs{x}}_{i},\bs{d}_i}\cup\{\sum_{i=1}^n\bs{d}_i\leq r, \mathcal{C}_\bs{in}^{\text{lb}}\leq \hat{\bs{x}}_{i}\leq \mathcal{C}_\bs{in}^{\text{ub}}\}$, where $\bs{d}$ is vector of fresh Boolean variables and for every $i \in [n]$, $\Psi_{g_i,\hat{\bs{x}}_{i},\bs{d}_i}$ is the encoding
  of the piecewise constant function $g_i$:
\begin{center}
  $g(\hat{\bs{x}}_{i}) =
\begin{cases}
  1, &  \text{if } \hat{\bs{x}}_{i}\in [\mathcal{C}_\bs{in}^{\text{lb}}, \hat{\bs{u}}_i);\\
  0, &  \text{if }  \hat{\bs{x}}_{i} \in [\hat{\bs{u}}_i, \hat{\bs{u}}_i+1);\\
  1, &  \text{if } \hat{\bs{x}}_{i} \in [\hat{\bs{u}}_i+1,\mathcal{C}_\bs{in}^{\text{ub}}).
\end{cases}$
\end{center}
Intuitively, for every $ \hat{\bs{x}}_{i}$ such that $\mathcal{C}_\bs{in}^{\text{lb}}\leq \hat{\bs{x}}_{i}\leq \mathcal{C}_\bs{in}^{\text{ub}}$, we have: $\Psi_{g_i,\hat{\bs{x}}_{i},\bs{d}_i}$ holds iff $g(\hat{\bs{x}}_{i})=\bs{d}_i$, while
$\bs{d}_i=1$ iff $\hat{\bs{u}}_i\neq \hat{\bs{x}}_{i}$, thus $\Phi_{\hat{\bs{u}},r}^0$ holds
iff the number of indices $i$ such that $\hat{\bs{u}}_i\neq \hat{\bs{x}}_{i}$ and $\mathcal{C}_\bs{in}^{\text{lb}}\leq \hat{\bs{x}}_{i}\leq \mathcal{C}_\bs{in}^{\text{ub}}$ is at most $r$.
\item   $\Phi_{\hat{\bs{u}},r}^1=\left\{
    -\bs{d}_i \leq\hat{\bs{x}}_i-\hat{\bs{u}}_i  \leq  \bs{d}_i, \mathcal{C}_\bs{in}^{\text{lb}}\leq \hat{\bs{x}}_{i}\leq \mathcal{C}_\bs{in}^{\text{ub}},
   \sum_{j=1}^{n}\bs{d}_j \le r
 \mid \ i\in [n]
  \right\}$
Clearly, $-\bs{d}_i \leq\hat{\bs{x}}_i-\hat{\bs{u}}_i  \leq  \bs{d}_i $ ensures
that $\bs{d}_i\geq |\hat{\bs{x}}_i-\hat{\bs{u}}_i|$. Together with $\sum_{i=j}^{n}\bs{d}_j \le r$, we get that $\Phi_{\hat{\bs{u}},r}^1$ holds
iff  $\hat{\bs{x}}\in\widehat{R}_1(\hat{\bs{u}},r)$.
  \item    $\Phi_{\hat{\bs{u}},r}^2=\big\{ \sum_{i=1}^n (\hat{\bs{x}}_i-\hat{\bs{u}}_i)^2\le r^2,  \mathcal{C}_\bs{in}^{\text{lb}}\leq \hat{\bs{x}}_{i}\leq \mathcal{C}_\bs{in}^{\text{ub}}\big\}$;
  \item  $\Phi_{\hat{\bs{u}},r}^\infty=\big\{ \max(\hat{\bs{u}}_i-r,\mathcal{C}_\bs{in}^{\text{lb}}) \leq  \hat{\bs{x}}_{i} \leq \min(\hat{\bs{u}}_i+r, \mathcal{C}_\bs{in}^{\text{ub}})\mid i\in [n]\big\}$.
  \end{itemize}

We can observe that $\Phi_{\hat{\bs{u}},r}^0$, $\Phi_{\hat{\bs{u}},r}^1$ and $\Phi_{\hat{\bs{u}},r}^\infty$ are integer linear constraints,
while $\Phi_{\hat{\bs{u}},r}^2$ is not. Although the quadratics in $\Phi_{\hat{\bs{u}},r}^2$ could be transformed into
integer linear constraints using bit-wise computations according to $\mathcal{C}_\bs{in}^{\text{lb}}\leq \hat{\bs{x}}_{i}\leq \mathcal{C}_\bs{in}^{\text{ub}}$, it is non-trivial.
 In fact, modern ILP-solvers (e.g., Gurobi)
support integer quadratic constraints.

\begin{example}
Consider the QNN $\widehat{\mb}_{e}$ given in Example~\ref{examp:qnn} and an input $\hat{\bs{u}}=(10,2)$.
We have: $\mathcal{C}_\bs{in}^{\text{lb}}=0$, $\mathcal{C}_\bs{in}^{\text{ub}}=63$ and  $\Phi_{\hat{\bs{u}},4}^\infty=\{6\leq \hat{\bs{x}}_1  \le 14,0\leq\hat{\bs{x}}_2 \le 6 \}$.

\end{example}

\subsection{Encoding Robustness Properties}
Recall that a QNN $\widehat{\mb}$ is robust w.r.t. an input region $\widehat{R}_p(\hat{\bs{u}},r)$ 
if all the input samples from $\widehat{R}_p(\hat{\bs{u}},r)$ have the same outputs (the same classes for classification tasks) as the input $\hat{\bs{u}}$. 
Thus, it suffices to check if $\widehat{\mb}(\hat{\bs{x}})\neq\widehat{\mb}(\hat{\bs{u}})$ for some $\hat{\bs{x}}\in\widehat{R}_p(\hat{\bs{u}},r)$
which is reduced to the solving of integer
linear constraints as follows. 
Suppose $\bs{y}=\widehat{\mb}(\hat{\bs{u}})$
and $\bs{y}^d=\widehat{\mb}(\hat{\bs{x}})$.


We introduce a set of Boolean variables $\{v^d_{i,0},v^d_{i,1}\mid i\in[m]\}$ such that for all $i\in[m]$,
the following constraints hold:
\begin{center}
$\hat{\bs{y}}^d_i > \bs{y}_i \Leftrightarrow v^d_{i,0}=1$ and $\hat{\bs{y}}^d_i < \bs{y}_i \Leftrightarrow v^d_{i,1}=1$.
\end{center}
%
The constraints $\hat{\bs{y}}^d_i > \bs{y}_i \Leftrightarrow v^d_{i,0}=1$ 
and $\hat{\bs{y}}^d_i < \bs{y}_i \Leftrightarrow v^d_{i,1}=1$ can then be encoded by the set $\Theta_{i}$ 
of integer linear constrains with extremely large integer number $\textbf{M}$:
\begin{center}
$\Theta_{i}= \left\{
\begin{array}{c}
\hat{\bs{y}}^d_i \ge  \bs{y}_i +1+\textbf{M}(v^d_{i,0}-1), \quad
  \hat{\bs{y}}^d_i <  \bs{y}_i +\textbf{M}v^d_{i,0},\\
 \bs{y}_i \ge  \hat{\bs{y}}^d_i +1+\textbf{M}(v^d_{i,1}-1), \quad
   \bs{y}_i < \hat{\bs{y}}^d_i +\textbf{M}v^d_{i,1}
\end{array} \right\}$
\end{center}
Intuitively, $\Theta_{i}$  ensures that $\hat{\bs{y}}^d_i \neq \bs{y}_i$ iff $v^d_{i,0}+v^d_{i,1} \ge 1$.

As a result, $\bs{y}^d\neq \bs{y}$, i.e., $\widehat{\mb}(\hat{\bs{x}})\neq\widehat{\mb}(\hat{\bs{u}})$,
iff the set $\Theta_{\hat{\bs{u}}}$ of integer linear constrains $\bigcup_i \Theta_{i} \cup \{\sum_{i\in[m]}v^d_{i,0}+v^d_{i,1}\ge 1\}$ holds.

For classification tasks, suppose $\widehat{\mb}^c(\hat{\bs{u}})=g$.
We introduce a set of Boolean variables $\{v^d_{i,g}\mid i\in[m]\}$ such that for all $i\in[m]$, 
the following constraints hold:
\begin{center}
  \begin{itemize}[leftmargin=*]
    \item If $i < g$, then  $\hat{\bs{y}}^d_i \ge \hat{\bs{y}}^d_g \Leftrightarrow v^d_{i,g}=1$ which can be encoded as
   $\Theta^c_{i,0}=\big\{\hat{\bs{y}}^d_i \ge  \hat{\bs{y}}^d_g + \textbf{M}(v^d_{i,g}-1), 
  \hat{\bs{y}}^d_i <  \hat{\bs{y}}^d_g + \textbf{M}v^d_{i,g}\big\}$;
    \item If $i > g$, then $\hat{\bs{y}}^d_i > \hat{\bs{y}}^d_g \Leftrightarrow v^d_{i,g}=1$ which can be encoded as
   $\Theta^c_{i,1}=\big\{  \hat{\bs{y}}^d_i \ge \hat{\bs{y}}^d_g +1+\textbf{M}(v^d_{i,g}-1),
  \hat{\bs{y}}^d_i < \hat{\bs{y}}^d_g +\textbf{M}v^d_{i,g}\big\}$.
  \end{itemize} 
\end{center} 
Intuitively, $\Theta^c_{i,0}$ (resp., $\Theta^c_{i,1}$) ensures that the $i$-th entry
of the output $\hat{\bs{y}}^d$  for $i< g$ (resp. $i>g$) is no less than (resp. larger than) the $g$-th entry iff $v^d_{i,g}=1$. 

As a result, $\widehat{\mb}^c(\hat{\bs{u}})=g\neq \widehat{\mb}^c(\hat{\bs{x}})$
iff the set $\Theta^c_{\hat{\bs{u}}}$ of integer linear constrains $\bigcup_{i<g} \Theta^c_{i,0} \cup \bigcup_{i>g} \Theta^c_{i,1}  \cup \{\sum_{i\in[m],i\neq g}v^d_{i,g}\ge 1\}$ holds.

We remark that though we focus on robustness properties of QNNs, any properties that
can be expressed in integer linear constraints could be verified with our approach.

\begin{example}
    Consider the QNN $\widehat{\mb}_{e}$ with an input $\hat{\bs{u}}=(10,2)$ in Example~\ref{examp:qnn}. Clearly, we have: $\widehat{\mb}_{e}^c(\hat{\bs{u}})=2$,
    and $\widehat{\mb}_{e}^c(\hat{\bs{x}})\neq 2$ iff $\Theta^c_{\hat{\bs{u}}}=\{\hat{\bs{z}}_1 \ge \hat{\bs{z}}_2 +\textbf{M}(v^d_{1,2}-1),\hat{\bs{z}}_1 < \hat{\bs{z}}_2 +\textbf{M}v^d_{1,2}, v^d_{1,2}\ge 1 \}$
    holds.
\end{example}
\vspace*{-2mm}

\begin{theorem}
    Given a QNN $\widehat{\mb}$ with $s$ neurons and quantization configuration $\mathcal{C}_\bs{out}$ for outputs of each non-input layer,
    we can reduce the robustness verification problem of $\widehat{\mb}$ into the solving of integer linear constraints with at most $\mathcal{O}(s)$ constraints and $\mathcal{O}(s\cdot ( \mathcal{C}_\bs{out}^{\text{ub}}-\mathcal{C}_\bs{out}^{\text{lb}}+2))$ integer and Boolean variables. \hfill $\qed$
\end{theorem}

\subsection{Constraint Simplification} \label{subsec:implDetail}

%
Recall that a Boolean variable is introduced for each interval when encoding piecewise constant functions (cf. Section~\ref{sec:pwcf2ilp}).
We observe that some intervals are guaranteed to never be taken for a given input region, thus their Boolean variables can be avoided.
To identify such intervals and thus reduce the size of constraints, we leverage interval analysis (IA)~\cite{WPWYJ18,TXT18} to soundly approximate the output ranges of neurons.
The estimated intervals can further be used to remove inactive neurons in non-input layers,
where the output of an inactive neuron is replaced by $\text{lb}$ if the estimated upper-bound is no greater than $\text{lb}$ ($\text{lb}$ is $\mathcal{C}_\bs{out}^{\text{lb}}$
for output layer and $0$ for hidden layers) or by $\mathcal{C}_\bs{out}^{\text{ub}}$ if the estimated lower-bound is no less than $\mathcal{C}_\bs{out}^{\text{ub}}$.

More specifically, given an input region $\widehat{R}_p(\hat{\bs{u}},r)$, we first compute the intervals $[\hat{\bs{x}}_i^\text{lb},\hat{\bs{x}}_i^\text{ub}]$ as the output ranges of the neurons $i \in[n]$ in the input layer
as follows:
\begin{itemize}[leftmargin=*]\setlength{\parskip}{2pt}
  \item For $\widehat{R}_0(\hat{\bs{u}},r)$: $\left\{\begin{array}{l}
                                                \hat{\bs{x}}_i^\text{lb}=\mathcal{C}_\bs{in}^\text{lb}, \smallskip\\
                                                \hat{\bs{x}}_i^\text{ub}=\mathcal{C}_\bs{in}^\text{ub},
                                              \end{array}\right.$ if $r\ge 1$;
                                              otherwise
                                              $\left\{\begin{array}{l}
                                                \hat{\bs{x}}_i^\text{lb}=\hat{\bs{u}}_i, \smallskip\\
                                                \hat{\bs{x}}_i^\text{ub}=\hat{\bs{u}}_i.
                                              \end{array}\right.$
  %
  \item For $\widehat{R}_1(\hat{\bs{u}},r)$ or $\widehat{R}_\infty(\hat{\bs{u}},r)$:
  $\left\{\begin{array}{l}
                                                \hat{\bs{x}}_i^\text{lb}=\text{clamp}(\hat{\bs{u}}_j-r,\mathcal{C}_\bs{in}^{\text{lb}},\mathcal{C}_\bs{in}^{\text{ub}}), \smallskip\\
                                                \hat{\bs{x}}_i^\text{ub}=\text{clamp}(\hat{\bs{u}}_j+r,\mathcal{C}_\bs{in}^{\text{lb}},\mathcal{C}_\bs{in}^{\text{ub}}).
                                              \end{array}\right.$
  \item For $\widehat{R}_2(\hat{\bs{u}},r)$: $\left\{\begin{array}{l}
                                                \hat{\bs{x}}_i^\text{lb}=\text{clamp}(\hat{\bs{u}}_j-\lfloor\sqrt{r}\rfloor,\mathcal{C}_\bs{in}^{\text{lb}},\mathcal{C}_\bs{in}^{\text{ub}}), \smallskip\\
                                                \hat{\bs{x}}_i^\text{ub}=\text{clamp}(\hat{\bs{u}}_j+\lfloor\sqrt{r}\rfloor,\mathcal{C}_\bs{in}^{\text{lb}},\mathcal{C}_\bs{in}^{\text{ub}}).
                                              \end{array}\right.$

\end{itemize}

We then soundly approximate the output ranges of other neurons by propagating the intervals through the network from the input layer to the output layer in the standard way~\cite{moore2009introduction}.
By doing so, we obtain an output range for each neuron.


\begin{figure}[t]
	\begin{center}
		\subfigure[Intervals]{
			\label{fig:QNN_IA_Fig}
			\includegraphics[width=0.205\textwidth]{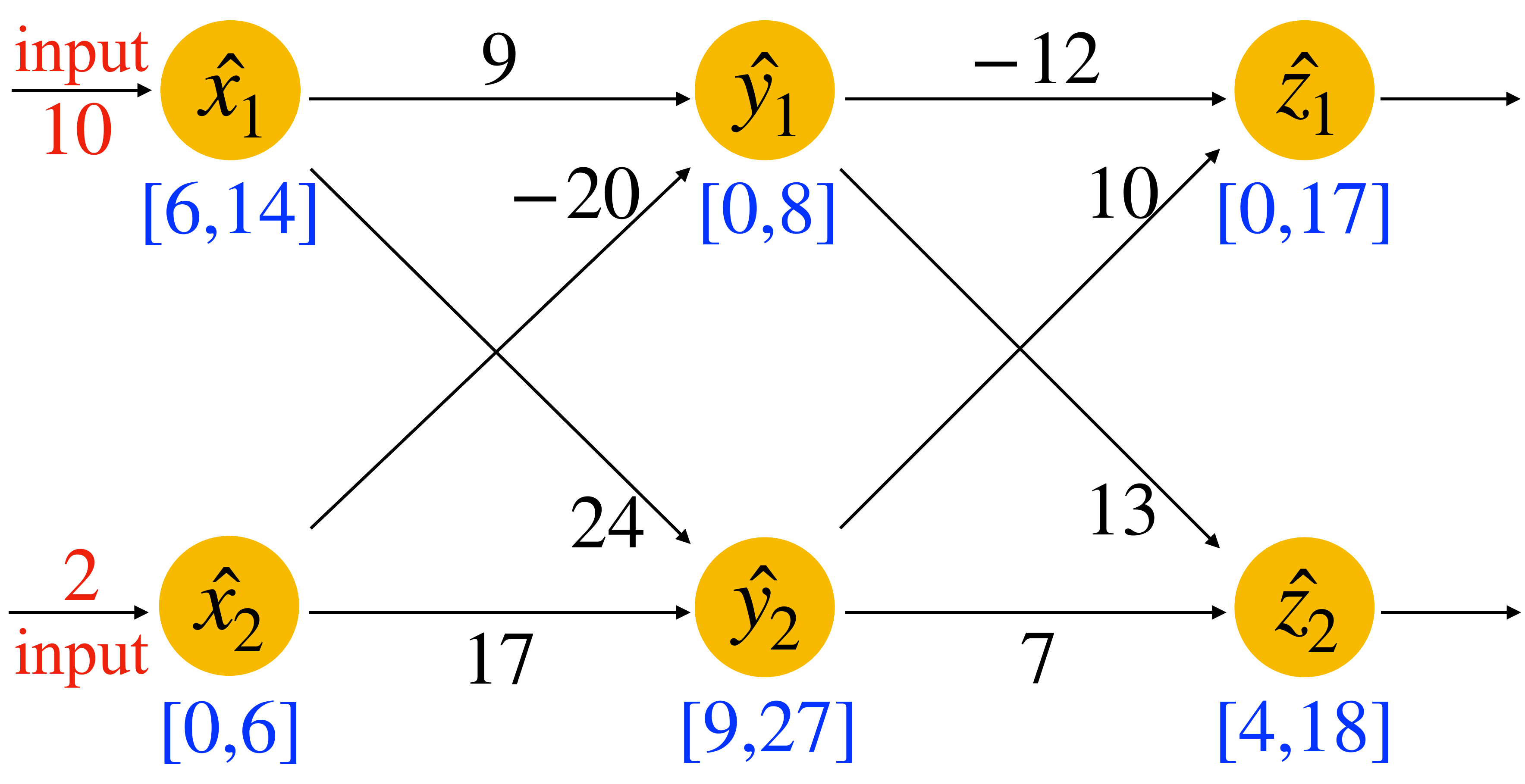}
		}
		\subfigure[Computation of intervals]{
			\label{fig:QNN_IA_Num}
			\includegraphics[width=0.245\textwidth]{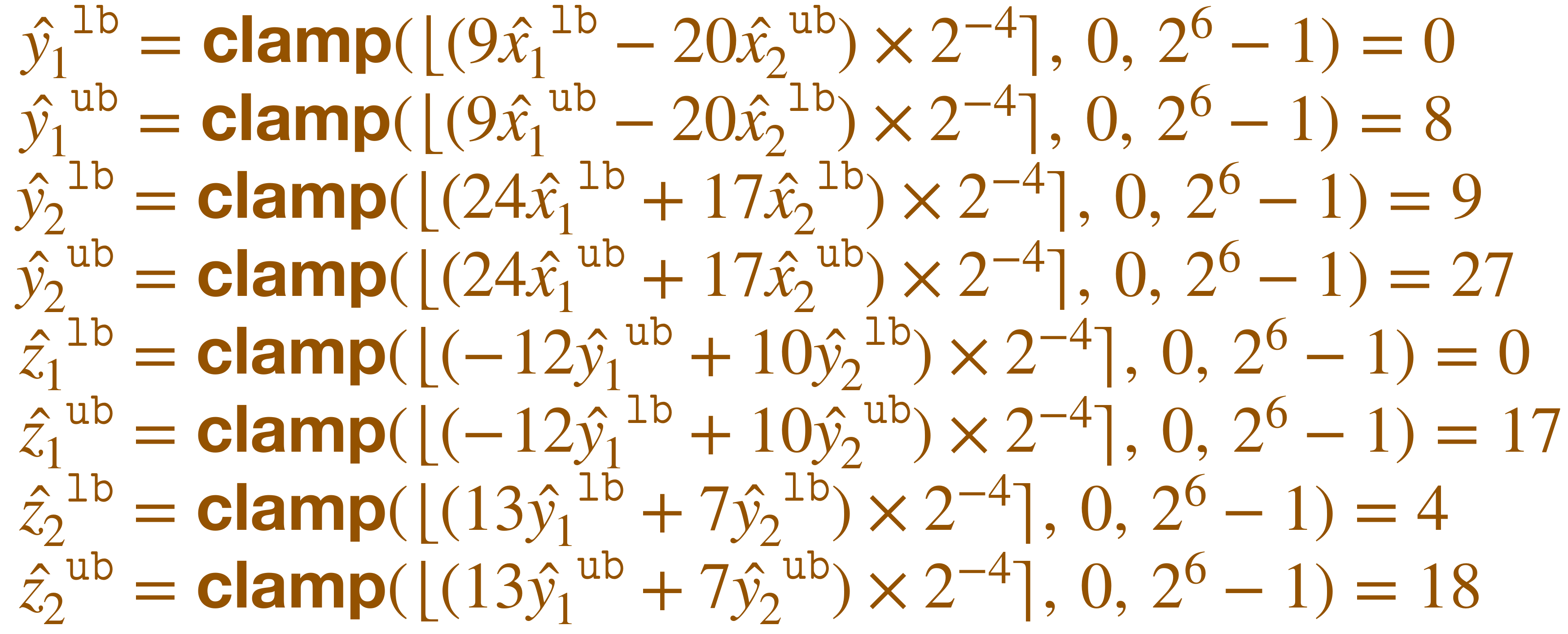}
		}
    \vspace{-3mm}
		\caption{Interval analysis on the QNN $\widehat{\mb}_{e}$.}
    \vspace{-4mm}
		\label{fig:propagation}
	\end{center}
\end{figure}

\begin{figure}[t]
	\centering
	\includegraphics[width=.4\textwidth]{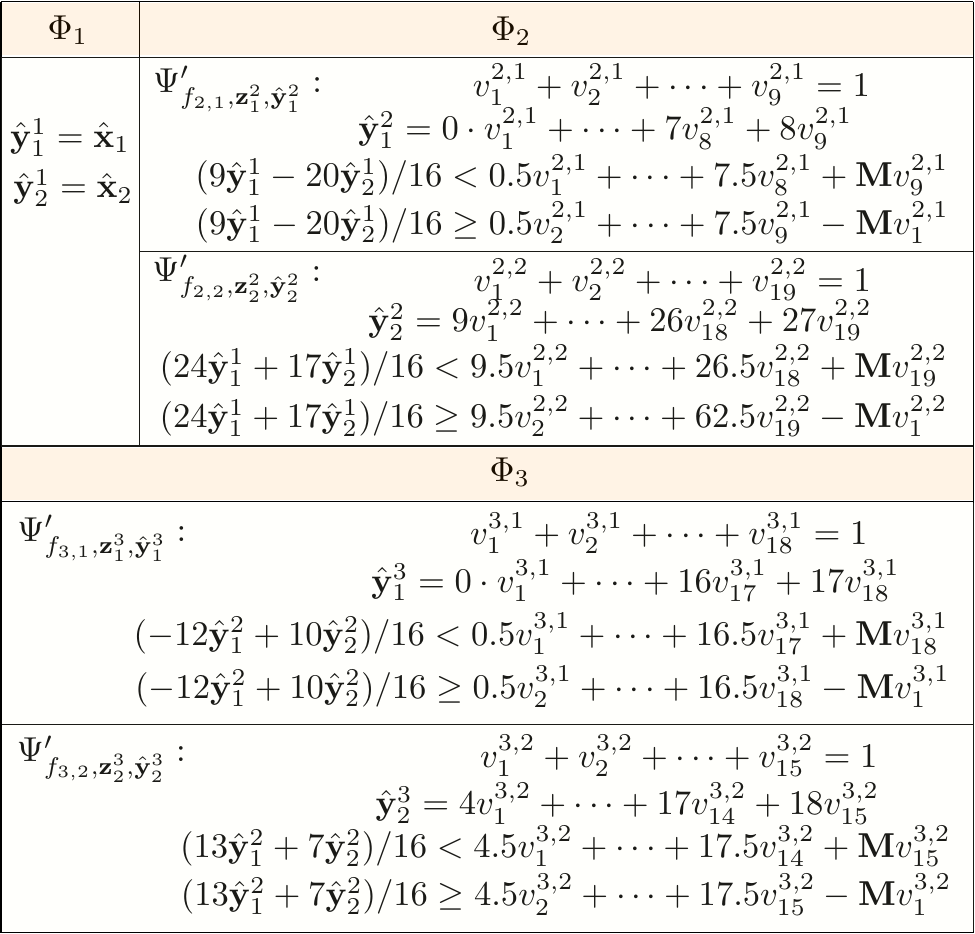}
  \vspace{-2mm}	
  \caption{Simplified encoding of $\widehat{\mb}_{e}$.}\label{fig:qnnILP_IA_demo}
  \vspace*{-3mm}
\end{figure}

\begin{example}\label{examp:qnn_IA}
  Consider the QNN $\widehat{\mb}_{e}$ given in Example~\ref{examp:qnn} with the input region $\widehat{R}_\infty(\hat{\bs{u}},r)$ where $\hat{\bs{u}}=(10,2)$ and $r=4$.
We first get the intervals $\hat{\bs{x}}_1\in[6,14]$ and $\hat{\bs{x}}_2\in[0,6]$ for the input layer.
As shown in Figure~\ref{fig:propagation}, we then compute the upper and  lower bounds for the neurons $\hat{\bs{y}}_1,\hat{\bs{y}}_2$ using the upper and lower bounds of $\hat{\bs{x}}_1,\hat{\bs{x}}_2$, which are later used to compute the upper and  lower bounds of the neurons $\hat{\bs{z}}_1,\hat{\bs{z}}_2$.
For instance, the lower bound $\hat{\bs{y}}_1^\text{lb}$ of $\hat{\bs{y}}_1$ is computed by the lower bound of $\hat{\bs{x}}_1$ and the upper bound of $\hat{\bs{x}}_2$ (the first equation in Figure~\ref{fig:QNN_IA_Num}),
because the clamp function increases monotonically, the weight between $\hat{\bs{x}}_1$ and $\hat{\bs{y}}_1$ is positive and the weight between $\hat{\bs{x}}_2$ and $\hat{\bs{y}}_1$ is negative.
The other bounds are computed similarly.
Based on the estimated intervals, the simplified encoding of $\widehat{\mb}_{e}$ is shown in Figure~\ref{fig:qnnILP_IA_demo},
which has fewer Boolean variables and smaller constraints than the ones given in Figure~\ref{fig:qnnILPdemo}.
\end{example}


\section{Computing Maximum Robustness Radius} \label{sec:maxR}
An straightforward approach to compute the maximum robustness radius (MRR) of a given input $\hat{\bs{u}}$
 is to transform the problem to an optimization problem based on our encoding with an additional objective function to maximize the attack radius, and solve it directly using an ILP solver.
However, this optimization problem is computationally intensive in general. As a consequence, our preliminary results show that this approach is only applicable to small-scale networks (cf. our website~\cite{website}).

\begin{algorithm}[t]
\small	\SetAlgoNoLine
\SetKwProg{myproc}{Proc}{}{}
\myproc{{\sc Main}{$({\tt QNN}: \widehat{\mb},\ {\tt Input}: \hat{\bs{u}}, \ {\tt StartR}: r, \ {\tt Step}: q,  \ {\tt Norm}: p)$}}{
  \If{{\sc VerifyR}$(\widehat{\mb},\hat{\bs{u}},1,p)$={\tt NonRobust}}   {
    \Return{$0$} \tcp*{$\widehat{\mb}$ is not robust w.r.t. $\widehat{R}_p(\hat{\bs{u}},1)$}
  }
  \Else{
    $r^\text{lb}$, $r^\text{ub}\leftarrow$  {\sc GetRange}($\widehat{\mb}, \ \hat{\bs{u}}, p, 1, r, q$) \;
	\Return{ {\sc GetMRR}$(\widehat{\mb}, \hat{\bs{u}}, p, r^\emph{lb}, r^\emph{ub})$\; }
   }
}
\smallskip

\myproc{{\sc GetRange}{$\left(\begin{array}{c}
                          {\tt QNN}: \widehat{\mb},\ {\tt Input}: \hat{\bs{u}}, \ {\tt Norm}: p,  \\
                          {\tt MinR}: r^\emph{lb}, \ {\tt MaxR}: r^\emph{ub}, \ {\tt Step}: q
                        \end{array}\right)$}}{
		\If{{\sc VerifyR}$(\widehat{\mb},\hat{\bs{u}},r^\emph{ub},p)$={\tt Robust}}{
			$r^\text{lb},r^\text{ub} \leftarrow r^\text{ub},  r^\text{ub}+q$ \;
			\Return{{\sc GetRange}$(\widehat{\mb}, \ \hat{\bs{u}}, \ p, r^\emph{lb}, \ r^\emph{ub})$\;}
		}
		\Return{$r^\emph{lb}$, $r^\emph{ub}$\;}
		
	}
	\smallskip
	\myproc{{\sc GetMRR}{$\left(\begin{array}{c}
                          {\tt QNN}: \widehat{\mb},\ {\tt Input}: \hat{\bs{u}}, \ {\tt Norm}: p,  \\
                          {\tt MinR}: r^\emph{lb}, \ {\tt MaxR}: r^\emph{ub}
                        \end{array}\right)$}}{
		\lIf{$r^\emph{ub}= r^\emph{lb}+1$}{
			\Return{$r^\text{lb}$}
		}
		\Else{
			$r' \leftarrow r^\text{lb} + (r^\text{ub} -r^\text{lb} )//2$ \;
			\If{{\sc VerifyR}$(\widehat{\mb},\hat{\bs{u}},r',p)$={\tt NonRobust}}{
				\Return{{\sc GetMRR}$(\widehat{\mb}, \ \hat{\bs{u}},  \ p, \ r^\emph{lb}, \ r')$\;}
			}
			\lElse{
				\Return{{\sc GetMRR}$(\widehat{\mb}, \ \hat{\bs{u}},  \ p, \  r', \ r^\emph{ub})$}
			}
		}
	}
	\caption{Computing MRR.}
	\label{alg:maxR}
\end{algorithm}

We present another approach which iteratively searches in a binary manner a potential radius $r$ for each input $\hat{\bs{u}}$ and check if the QNN is robust or not w.r.t.\ the input region $\widehat{R}_p(\hat{\bs{u}},r)$ by invoking our robustness verification procedure, named {\sc VerifyR}$(\widehat{\mb},\hat{\bs{u}},r,p)$.
Our approach is described in Algorithm~\ref{alg:maxR}. It works as follows.

Given a QNN $\widehat{\mb}$ and an input $\hat{\bs{u}}$, it first checks if $\widehat{\mb}$ is robust w.r.t.  $\widehat{R}_p(\hat{\bs{u}},1)$, i.e, the attack radius is $1$, by invoking {\sc VerifyR}$(\widehat{\mb},\hat{\bs{u}},1,p)$.
If {\sc VerifyR}$(\widehat{\mb},\hat{\bs{u}},1,p)$ returns {\tt NonRobust}, MRR is $0$. 
Otherwise, $\widehat{\mb}$ is robust w.r.t.  $\widehat{R}_p(\hat{\bs{u}},1)$. We then
invoke the procedure {\sc GetRange} to quickly determine a range $[r^\text{lb},r^\text{ub}]$ such that
$\widehat{\mb}$ is robust w.r.t. $\widehat{R}_p(\hat{\bs{u}},r^\text{lb})$ but is not robust w.r.t., $\widehat{R}_p(\hat{\bs{u}},r^\text{ub})$,
where the initial value of $r^\text{lb}$ is $1$ and the initial value of $r^\text{ub}$ is $r$ provided by users.
Next, it invokes the procedure {\sc GetMRR}($\widehat{\mb},\hat{\bs{u}},p,r^\text{lb},r^\text{ub}$)
to search the MRR between $r^\text{lb}$ and $r^\text{ub}$ in the standard binary search manner.

The procedure {\sc GetRange} receives a range $[r^\text{lb},r^\text{ub}]$ and tries to quickly update the range such that
$\widehat{\mb}$ is robust w.r.t. $\widehat{R}_p(\hat{\bs{u}},r^\text{lb})$ but is not robust w.r.t., $\widehat{R}_p(\hat{\bs{u}},r^\text{ub})$.
It first checks if $\widehat{\mb}$ is robust w.r.t. $\widehat{R}_p(\hat{\bs{u}},r^\text{ub})$.
If it is the case, the range $[r^\text{lb},r^\text{ub}]$ is updated to $[r^\text{ub},r^\text{ub}+q]$
on which {\sc GetRange} is recursively invoked,
where $q$ is a given range size used as the search step for searching the desired range efficiently.
Otherwise, the current range $[r^\text{lb},r^\text{ub}]$  is returned.
We note that it is important to select an appropriate radius $r$ as the initial $r^\text{ub}$ and the step $q$ to compute the desired range efficiently.
We remark that such a binary-search-based method can be used to find the maximum robustness radius for DNNs only if the attack radii are restricted to fixed-point values~\cite{li2022towards}.

\section{Evaluation}\label{sec:evaluation}

We have implemented our approach as a tool \tool, where the ILP-solver Gurobi \cite{Gurobi} is utilized as the constraint solving engine.
Gurobi does not support strict inequalities (e.g., less-than) which are used in our encoding of piecewise constant functions.
To address this issue, we reformulate each integer linear constraint $g(x_1,\cdots, x_k)< 0$  into  $g(x_1,\cdots, x_k) + \epsilon \le 0$ where $\epsilon >0$ is a constant smaller than the precision of $g(x_1,\cdots, x_k)$.
For example, the integer linear constraint $(9\hat{\bs{y}}^1_1-20\hat{\bs{y}}^1_2)/16 < 0.5 v_1^{2,1}+ \cdots+ 7.5 v_{8}^{2,1}+\textbf{M} v_9^{2,1}$ of $\widehat{\bs N}_e$ (cf. Figure~\ref{fig:qnnILP_IA_demo})
is written as $(9\hat{\bs{y}}^1_1-20\hat{\bs{y}}^1_2)/16 -( 0.5 v_1^{2,1}+ \cdots+ 7.5 v_{8}^{2,1}+\textbf{M} v_9^{2,1})+ \epsilon \le 0$.
As $\{\hat{\bs{y}}_1^1,\hat{\bs{y}}_1^2\}$ and $\{v_1^{2,1},\ldots,v_9^{2,1}\}$ are integer/Boolean variables, the precision of $(9\hat{\bs{y}}^1_1-20\hat{\bs{y}}^1_2)/16 - ( 0.5 v_1^{2,1}+ \cdots+ 7.5 v_{8}^{2,1}+\textbf{M} v_9^{2,1} )$ is $1/16=0.0625$.
Thus, $\epsilon$ can be any value s.t. $0< \epsilon < 0.0625$.

\begin{table}[t]
	\centering
	\caption{QNN benchmarks and accuracy, QNNs taken from~\cite{scaleQNN21} are marked by $\dag$.}
	\vspace*{-3mm}
	\label{tab:bench}\setlength{\tabcolsep}{5pt} 
	\scalebox{0.88}{
		\begin{tabular}{c|c|c|c|c|c}
			\toprule
			Dataset & ARCH & $Q=4$ & $Q=6$ & $Q=8$ & $Q=10$  \\ \midrule
			\rowcolor{gray!20}
			MNIST & P1 (784:64:10) & 96.80\%  & 97.08\%$^\dag$ & 97.10\% & 96.89\% \\
			
			MNIST&P2 (784:100:10) & 97.36\%& 97.58\% & 97.45\% & 97.35\%\\
			\rowcolor{gray!20}
			MNIST&P3 (784:64:32:10) & 96.91\% & 97.12\% & 97.01\% & 97.07\%\\
			
			F-MNIST& P4 (784:64:10) & -- &  85.60\%$^\dag$ & -- & -- \\
			\bottomrule
	\end{tabular}}
	\vspace*{-4mm}
\end{table}

We evaluate \tool to answer the following research questions:

\begin{enumerate}[label=\textbf{RQ\arabic*. },itemindent=*,leftmargin=*,itemsep=0pt]
    \item How effective is interval analysis for reducing the size of integer linear constraints and verification cost?
    \item How efficient and effective is \tool for robustness verification, compared over the state-of-the-art tools?
    \item How efficient and effective is \tool for computing MRR?
\end{enumerate}



\smallskip
\noindent{\bf Benchmarks.}
We use the quantization-aware training framework provided in~\cite{scaleQNN21} to train 11 QNNs using 3 architectures, the MNIST dataset~\cite{MNIST}, and  quantization configurations $\mathcal{C}_{\bs{in}}=\langle +, 8,8\rangle$, $\mathcal{C}_\bs{w}=\langle \pm , Q,Q-1\rangle$, $\mathcal{C}_\bs{b}=\langle \pm , Q,Q-2\rangle$, $\mathcal{C}_\bs{out}=\langle + , Q,Q-2\rangle$ for the hidden layers and $\mathcal{C}_\bs{out}=\langle \pm, Q,Q-2\rangle$ for the output layer for a given architecture, where $Q\in\{4,6,8,10\}$.
Together with 2 QNNs provided in~\cite{scaleQNN21}
 on which we compare with \cite{scaleQNN21}, we obtain 13 QNNs.
Details of the QNNs are shown in Table~\ref{tab:bench}.
Column 1 gives the dataset used for training.
Column 2 shows the name and architecture of the QNN, where $n_1:n_2:\cdots:n_d$ denotes that the QNN has $d$ layers, $n_1$ inputs and $n_d$ outputs,
along with $n_i$ neurons in each hidden layer $\ell_i$ for $2\le i\le d-1$.
Columns 3-6 list the accuracy of these QNNs.
Hereafter, we denote by P$x$-$y$ the QNN using the architecture P$x$ and quantization bit size $Q=y$.
We can observe that all the 12 QNNs trained on MNIST achieved more than 96.80\% accuracy.

The experiments were conducted on a 28-core machine with Intel Xeon E5-2690 2.6 GHz CPU and 251 GB main memory.
Note that, when using multiple threads for Gurobi, we set the number of threads to 28.

\subsection{RQ1: Effectiveness of Interval Analysis} \label{subsec:IA}
To answer {\bf RQ1}, we compare the number of Boolean variables/product terms in integer linear constraints and verification time for robustness verification of the QNN P1-6
by \tool with and without interval analysis.
We randomly select 100 input samples from the test set of MNIST that can be correctly predicted by the QNN P1-6.
The input regions are given by the $L_\infty$ norm with the attack radii $r\in\{1, 2, 4, 6, 10, 20, 30\}$ for each sample, resulting in 700 different input regions.
We encode the robustness verification tasks of the QNN  P1-6 w.r.t.\ all the 700 input regions with and without interval analysis, resulting a total of 1,400 verification tasks. 



\begin{figure}[t]
	\centering\setlength{\tabcolsep}{4pt}
	\includegraphics[width=.46\textwidth]{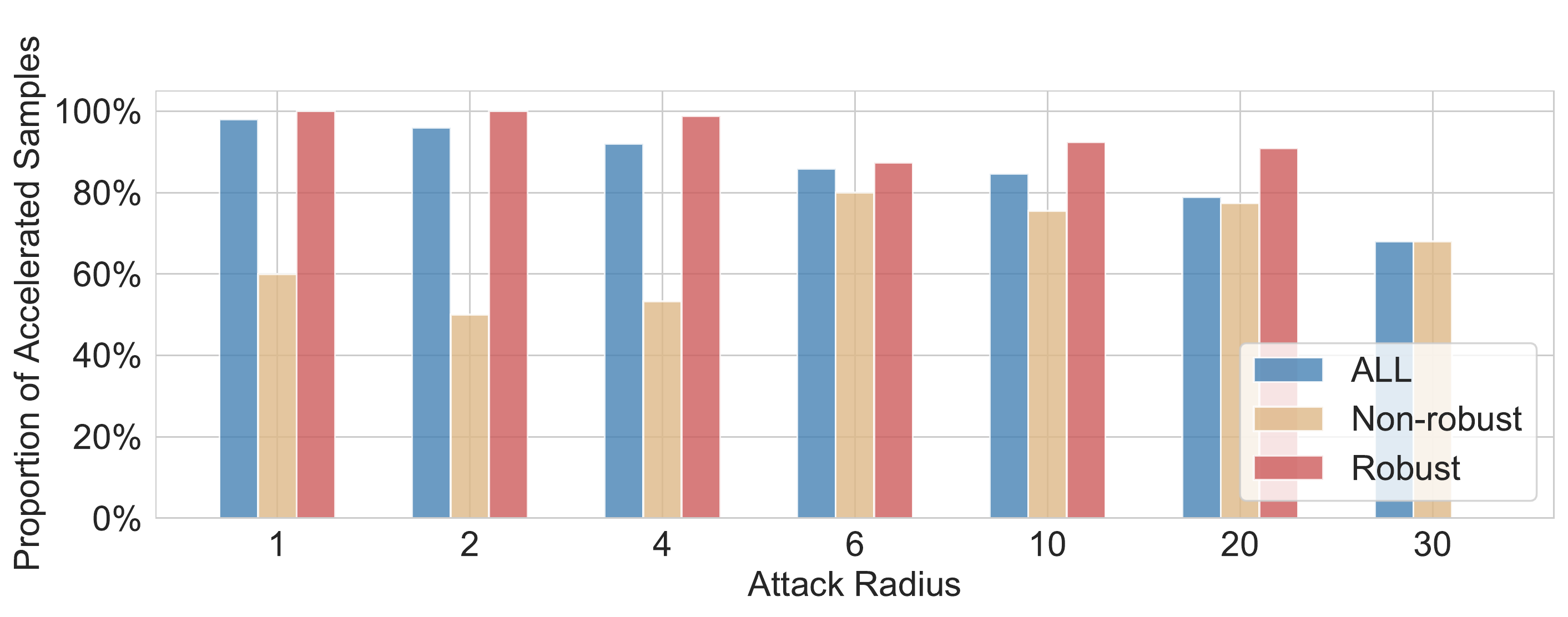}
	\scalebox{.86}{\begin{tabular}{c|c|c|c|c|c|c|c}
		\toprule
		Attack radius $r$ & 1 & 2 & 4 &6 &10 &20 &30 \\
		\midrule
		\rowcolor{gray!20}
		\makecell[c]{BVar/PTerms\\ Reduction Rate} & 84.6\% & 82.3\% & 76.9\% & 71.2\% & 59.6\% & 35.9\% & 22.1\%  \\
		Time (with IA)& 0.08 & 0.67 & 4.26 & 9.21 & 43.48 & 41.41 & 5.41  \\
		\rowcolor{gray!20}
		Time (without IA)& 0.23 &0.66 & 5.51 & 17.19 & 169.2 & 98.32 & 12.95  \\
		\bottomrule
	\end{tabular}}
	\vspace*{-2mm}
	\caption{Results of \tool with and without interval analysis.}\label{fig:ip_effec}
	\vspace*{-4mm}
\end{figure}

The results are given in Figure~\ref{fig:ip_effec} within 2 hours per task.
The histogram shows the distribution of the tasks that can be accelerated using interval analysis, where
the blue bars give the proportion of accelerated tasks among all the solved tasks,
and red (resp. yellow) bars give the proportion of accelerated tasks among all the solved tasks that are robust (resp. non-robust).
Note that no samples are robust under the attack radius 30.
The first row in the table shows the reduction rate of the number of Boolean variables (resp. product terms) in the integer linear constraints with interval analysis
and the other two rows show the average verification time (in seconds) of the solved tasks by \tool with and without interval analysis.

Overall, we can observe that the interval analysis can drastically reduce the number of Boolean variables and product terms in the integer linear constraints, thus reduce verification cost.
Unsurprisingly, we also observe that the reduction of the number of Boolean variables and product terms decreases with the increase of attack radius. It is because, with the increase of attack radius, the estimated interval becomes larger and thus
the number of Boolean variables/product terms that could be reduced becomes smaller.
Nevertheless, it could be quickly falsified when the attack radius is larger, thus
the interval analysis is able to significantly reduce the verification cost.
%
%
We note that the proportion of accelerated tasks among all the solved tasks that are robust (resp. non-robust) may increase
with attack radius, i.e., red (resp. yellow) bars, but it decreases among all the solved (i.e., blue bars).
It is because more and more input samples become non-robust with the increase of attack radius.

\begin{tcolorbox}[left=0mm,right=0mm,top=0mm,bottom=0mm]
	\textbf{Answer to RQ1:}
	Interval analysis is very effective in reducing the size of the integer linear constraints and verification cost. 
\end{tcolorbox}

\subsection{RQ2: \tool for Robustness Verification}

To answer {\bf RQ2}, we first compare the performance of \tool against the state-of-the-art verifier~\cite{smtQNN,scaleQNN21} for robustness verification
using QNNs of varied architectures but the same quantization bit size as~\cite{smtQNN,scaleQNN21} (i.e., P1-6, P2-6, P3-6 and P4-6).
We then evaluate the scalability of \tool on the 12 MNIST QNNs, i.e., P$x$-$y$ for $x\in\{1,2,3\}$ and $y\in\{4,6,8,10\}$.

\smallskip
\noindent{\bf Compared with the SMT-based verifier~\cite{smtQNN,scaleQNN21}.}
The input regions in this comparison are completely the same as~\cite{smtQNN,scaleQNN21}:
%
\begin{itemize}[leftmargin=*]
	\item For P1-6 and P2-6,  the attack radius $r$ is $1$ (resp. 2, 3 and 4) for the samples from the test set of MNIST with IDs 0$\sim$99 (resp. 100$\sim$199, 200$\sim$299 and 300$\sim$399).
After removing all the samples that cannot be correctly predicated by P1-6, there are 99, 99, 96 and 97 samples for $r=1,2,3,4$, respectively.
For P2-6, there are 98, 100, 96 and 99 samples predicted correctly.
	\item For P3-6, the attack radius $r$ is 1 (resp. 2, 3 and 4) for the samples from the test set of Fashion-MNIST with IDs 0$\sim$49 (resp. 100$\sim$149, 200$\sim$249 and 300$\sim$349).
After removing all the samples that cannot be correctly predicated by P3-6, there are 50, 49, 48 and 49 samples for $r=1,2,3,4$, respectively.
	\item For P4-6, the attack radius $r$ is 1 (resp. 2, 3 and 4) for the samples from the test set of Fashion-MNIST with IDs 0$\sim$99 (resp. 100$\sim$199, 200$\sim$249 and 300$\sim$349).
After removing all the samples that cannot be correctly predicated by P4-6, there are 87, 90, 43 and 44 samples for $r=1,2,3,4$, respectively.
\end{itemize}



The results are reported in Table~\ref{tab:cmpAI}.
We note that the SMT-based tool only supports single-thread~\cite{smtQNN,scaleQNN21} while
our tool \tool supports both single- and multiple-thread. Thus, we report results of \tool in
both single- and multiple-thread modes, where the multiple-thread mode uses 28 threads. Column (\#S) gives the number of verification tasks for each attack radius and each QNN. Columns (SR) show the success rate of verification within 2 hours per task.
Columns (Time) give the sum of verification time (in seconds) of all the solved tasks within 2 hours per task.
Note that, in Columns 7 and 9, we also show the sum of verification time $(n)$ of \tool for solving the verification tasks that are successfully solved by the SMT-based method~\cite{smtQNN,scaleQNN21} within 2 hours per task in the single- and multiple-thread modes, respectively.

We can observe that \tool solved significantly more verification tasks than the SMT-based method~\cite{smtQNN,scaleQNN21} within the same time limit (2 hours) in both the single- and multiple-thread modes.
In fact, \tool solved almost all the verification tasks for P1-6 and P4-6 within 2 hours, i.e.,
 only one verification task cannot be solved. It can be solved by neither \tool nor the SMT-based method~\cite{smtQNN,scaleQNN21} even using 24 hours, indicating it
is a very hard problem. For the verification tasks that can be solved by the SMT-based method~\cite{smtQNN,scaleQNN21},
\tool is two orders of magnitude faster than the state-of-the-art~\cite{smtQNN,scaleQNN21} in both the single- and multiple-thread modes.
Our conjecture is
that the SMT-based method~\cite{smtQNN,scaleQNN21} encodes the verification problem of QNNs using quantifier-free bit-vector (where the constants and bit-width  are given in binary representation)
and the resulting SMT constraints have a large number of If-Then-Else operations, both of which affect the runtime of the SMT-solver significantly~\cite{scaleQNN21}, although interval analysis is also
leveraged to reduce them. In contrast, we eliminate explicit If-Then-Else operations by introducing Boolean variables and the input/output of each neuron are encoded by integer variables.
On the other hand, the decision procedures of SMT-solver and ILP-solver are completely different, indicating
that the ILP-solver Gurobi is more suitable for solving constraints generated from the QNN verification problem.

We also observe that \tool with multiple threads performs better on harder tasks,
while \tool with a single thread performs better on easier tasks. For example, considering P4-6 with $r=1$.
For the 76 tasks that can be solved by SMT-based methods within 2 hours,
\tool takes 0.80 seconds using a single thread while 1.29 seconds using multiple threads.
However, for the rest 11 harder tasks that cannot be solved by SMT-based method, \tool with a single thread takes 992.6 seconds
while \tool with multiple threads only takes 48.06 seconds. In the following experiments, we use \tool with multiple threads by default.

\begin{table}[t]
	\centering
	\caption{\tool vs. the SMT-based verifier~\cite{smtQNN,scaleQNN21}.}
	\vspace*{-3mm}
	\label{tab:cmpAI}\setlength{\tabcolsep}{2pt} 
	 \scalebox{0.79}{\begin{tabular}{c|c|c|c|cc|cc|cc}
		\toprule
		\multirow{3}*{Dataset}& \multirow{3}*{QNN} &\multirow{3}*{r}&\multirow{3}*{\#S}& &&\multicolumn{4}{c}{\tool}\\
		\cline{7-10}
		~ & ~ & ~ & ~ & \multicolumn{2}{c|}{\multirow{-2}*{SMT-based}} & \multicolumn{2}{c|}{Single-thread} & \multicolumn{2}{c}{Multi-thread}\\
		\cline{5-10}

		~ & ~ & ~ & ~ & SR &Time(s) & SR &Time(s) & SR &Time(s)\\ \midrule
		\rowcolor{gray!20}
		\cellcolor{white} & \cellcolor{white} &1 & 99&  100\% & 14.78 & 100\% & 0.62 (0.62) & 100\% & 1.06 (1.06) \\
		
		 ~ &~ &2 & 99 & 93.9\%  & 3,388  & 100\% & 8.24 (1.55) &100\%  & 10.0 (2.04) \\
		\rowcolor{gray!20}
		\cellcolor{white} &\cellcolor{white}{\multirow{-2}*{P1-6}} &3& 96& 72.9\% & 352.6 & 100\% & 210.6 (0.86) & 100\%  & 78.09 (1.47)  \\
		
		~ & ~ &4& 97& 54.6\% &4,800 & 100\% & 779.7 (10.63) &100\%  & 792.8 (3.22) \\
		\hhline{~|---------}

		\rowcolor{gray!20}
		\cellcolor{white} & \cellcolor{white} &1 & 98 & 83.7\% & 3,074 & 100\% & 80.50 (53.59) & 100\% & 17.65 (4.27) \\
		
		\cellcolor{white} &~ & 2 & 100 & 22\% & 3,505 & 97\% & 357.9 (3.36) & 97\% & 194.5 (5.51)\\
		\rowcolor{gray!20}
		\cellcolor{white}{\multirow{-2}*{MNIST}} &\cellcolor{white}{\multirow{-2}*{P2-6}} &3& 96 & 4.17\% & 417.0 & 94.8\% & 683.7 (0.25) & 96.9\% & 1,952 (0.28)\\
		
		 & & 4 & 99 & 0\% & -- &80.8\% & 1,374 (--)& 87.9\% & 11,166 (--) \\
		 \hhline{~|---------}


		\rowcolor{gray!20}
		\cellcolor{white} &\cellcolor{white} &1 & 50& 42\% & 20,383 & 98\% & 4,936 (11.49) & 100\% & 869.9 (8.53)\\
		
		 ~ &~ &2 & 49 & 6.12\% & 3,759 & 85.7\% & 37,792 (38.69) & 95.9\% & 15,699 (36.54) \\
		\rowcolor{gray!20}
		\cellcolor{white} &\cellcolor{white}{\multirow{-2}*{P3-6}} &3& 48 & 0\% & -- & 43.8\% & 26,070 (--) & 50\% & 9,559 (--) \\
		
		\cellcolor{white} & &4& 49 & 0\% & -- & 20.4\% & 13,735 (--)& 24.5\% & 11,683 (--) \\ \midrule
		\rowcolor{gray!20}
		\cellcolor{white} & \cellcolor{white} &1 & 87 & 87.4\% & 120.6  & 100\% & 993.4 (0.80) & 100\% &  49.35 (1.29) \\
		
		 ~ & &2 & 90& 81.1\% & 4,491 & 100\% & 838.7 (21.98) & 100\% & 37.47 (3.19) \\
		\rowcolor{gray!20}
		\cellcolor{white}{\multirow{-2}*{F-MNIST}} &\cellcolor{white}{\multirow{-2}*{P4-6}} &3& 43&  60.4\%  &4,937 & 100\% & 324.5 (1.27)& 100\%  & 99.75 (2.57)  \\
		
		\cellcolor{white} & &4&44& 40.9\%   &136.2 &97.7\% &  987.6 (0.43)&97.7\% & 138.5 (0.66) \\
		\bottomrule
	\end{tabular}}
	\vspace*{-5mm}
\end{table}

\smallskip
\noindent
{\bf Scalability.} To understand the scalability of \tool better,
we evaluate \tool on the 12 MNIST QNNs, i.e., P$x$-$y$ for $x\in\{1,2,3\}$ and $y\in\{4,6,8,10\}$, with larger attack radii $r\in\{2,4,6,10,20,30\}$.
Following~\cite{smtQNN,scaleQNN21}, the input regions are designed as follows:
\begin{itemize}[leftmargin=*]
\item For P1-$y$ and P2-$y$, we select 100 samples with ID varied in the interval $[(r-1)\times100, r\times 100-1]$ from the test set of MNIST for each  $r$,
e.g., 100 samples with IDs $1900,\cdots, 1999$ for $r=20$, where
samples that cannot be correctly predicated by the QNN are removed, giving rise to at most 100 tasks for each $r$ and QNN;
	\item For P$3$-$y$, we select 50 samples with ID varied in the interval $[(r-1)\times 100, r\times 100-51]$ from the test set of MNIST for each $r$,
where samples that cannot be correctly predicated by the QNN are removed, resulting at most 50 tasks for each $r$.
\end{itemize}

\begin{table*}[t]
	\centering
	\caption{Success rate and verification time of \tool on MNIST dataset.}
	\vspace*{-2mm}
	\label{tab:scal_num}\setlength{\tabcolsep}{6pt} 
	\scalebox{0.85}{
		\begin{tabular}{c|c|ccc|ccc|ccc|ccc|ccc|ccc}
		\toprule
		 ~ & ~ &\multicolumn{3}{c|}{r=2}&\multicolumn{3}{c|}{r=4}&\multicolumn{3}{c|}{r=6}&\multicolumn{3}{c|}{r=10}&\multicolumn{3}{c|}{r=20}&\multicolumn{3}{c}{r=30}\\
		 \multirow{-2}*{ARCH} & \multirow{-2}*{$Q$} &\#S&SR&Time&\#S&SR&Time&\#S&SR &Time&\#S&SR&Time&\#S&SR&Time&\#S&SR&Time \\ \midrule
		\rowcolor{gray!20}
		\cellcolor{white} & 4  & 97 & 100\% &25.35& 96 & 100\% &34.68& 97 & 90.7\% &165.4 & 94 & 94.7\% &345.9& 97 & 99.0\% &53.21& 95 & 100\% &2.66\\
		
		\cellcolor{white} & 6  & 99 & 100\% &0.10& 97 & 100\% &8.17& 97 & 96.9\% & 12.60&97 & 94.8\% &204.0& 95 & 95.8\% &60.83& 93 & 100\%&9.40 \\
		\rowcolor{gray!20}
		\cellcolor{white}{\multirow{-2}*{P1}} & 8 & 100 & 100\% &0.02& 97 & 100\%&0.08 & 97 & 100\%&0.23 & 96 & 100\%&0.54 & 97 &100\%&0.82 & 96 & 87.65\%&90.47  \\
		
		\cellcolor{white} &10& 99 & 100\% &0.02& 98 & 100\%&0.11 & 96 & 100\%&0.35 & 95 & 100\%&0.92 & 93 &100\% &1.62& 93 &100\%&21.59 \\\midrule
		\rowcolor{gray!20}
		\cellcolor{white} & 4 &99& 100\%&17.3&98 &95.9\%&229.4 &97 & 72.2\% &379.0& 96&64.9\%&575.9 &95 &96.8\%&276.3& 96 &100\%&28.96\\
		
		\cellcolor{white} & 6 &100 &97\%&2.00 & 99& 87.9\%&128.3 & 100& 61\%&282.8 & 94 & 27.6\%&1,107& 96&68.8\%&574.4 &96 &91.7\%& 221.6\\
		\rowcolor{gray!20}
		\cellcolor{white}{\multirow{-2}*{P2}} & 8& 99 & 100\%&0.04 &98&100\%&0.26&99&100\%&0.48&95&100\%&0.79&93 &100\%&3.88 &95 &62.1\%&143.6   \\
		
		\cellcolor{white} &10& 99&100\%&0.03&97 &100\%&0.28 &97 &100\%&0.63 &96 &100\%&1.13 & 97&100\%&1.77& 96&91.7\%&31.50  \\\midrule
		\rowcolor{gray!20}
		\cellcolor{white} & 4 & 50&100\% &277.1& 49& 44.9\%&348.4 & 50 & 26\%&1,406& 49& 32.7\%&1,546& 48& 75\%&1,252& 46& 89.1\%&549.9 \\
		
		\cellcolor{white} & 6 & 49&95.9\% &334.0& 49 & 24.5\%&973.6 & 50 & 4\%&3,658& 49& 10.2\%&1,137& 47& 29.8\%&1,510& 47 &74.5\%&951.5  \\
		\rowcolor{gray!20}
		\cellcolor{white}{\multirow{-2}*{P3}} &8 &48&100\% &0.27& 47 &100\% &0.45& 50&100\%&0.51 & 48&100\%&2.99 &46 &100\%  &43.61& 47 & 74.5\% &513.0 \\
		
		\cellcolor{white} &10&49 &100\%&0.55&47 &100\%&0.94 &50 &100\%&1.13 &49 &100\%&1.57 &48 &100\%&345.7 &47 &100\%&568.6   \\
		\bottomrule
	\end{tabular}
	}
	\vspace*{-1mm}	
\end{table*}

\begin{figure}[t]
	\centering
	\includegraphics[width=.45\textwidth]{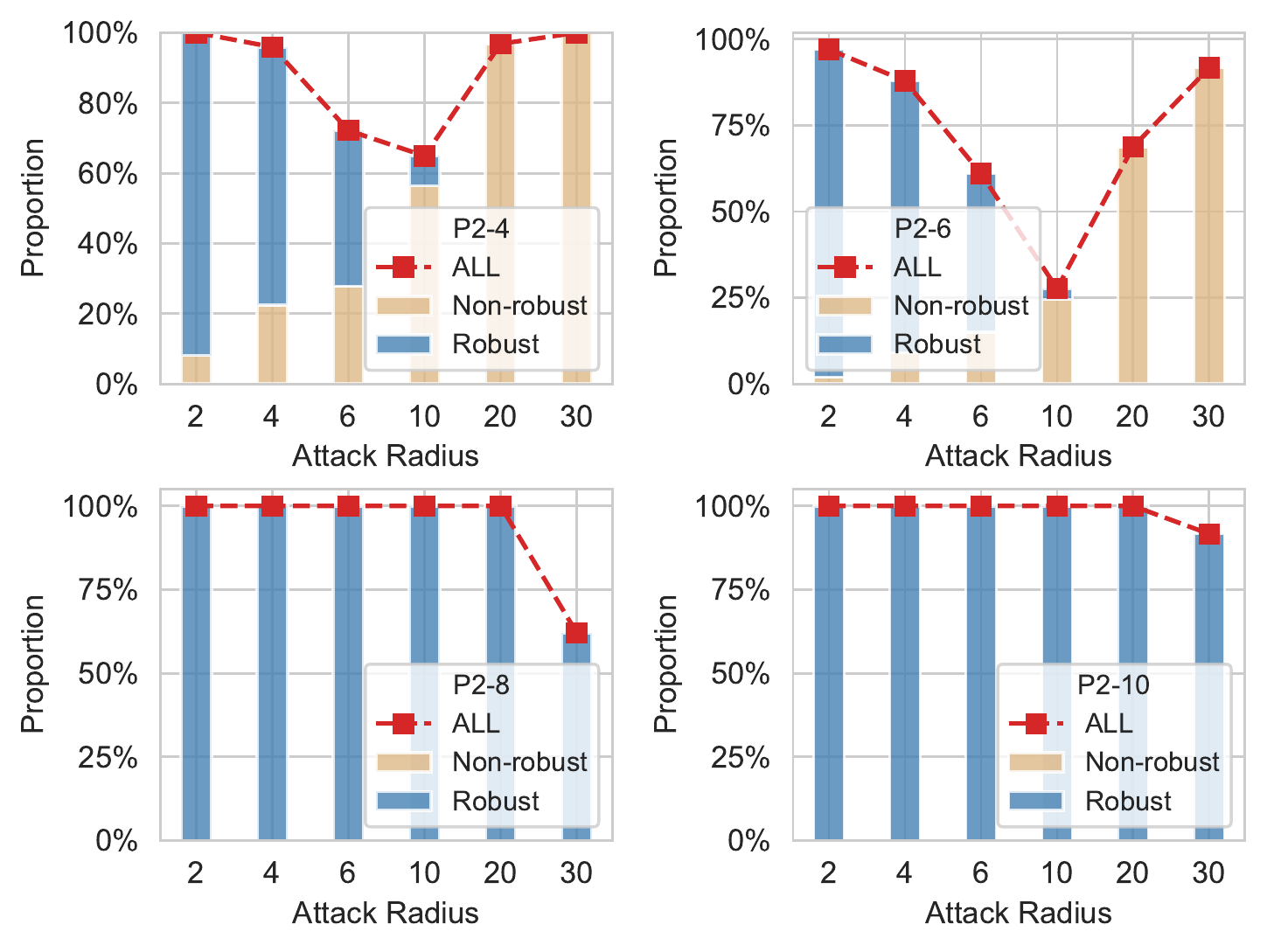}
	 \vspace*{-3mm}
	\caption{Distribution of (non-)robust samples under various attack radii $r$ and quantization bits $Q$.}
	\label{fig:dis_samples}
	\vspace*{-3mm}
\end{figure}
%

Table~\ref{tab:scal_num} shows the results of \tool within 2 hours per task, where
columns (\#S), (SR) and (Time) respectively give the number of verification tasks for each attack radius and each QNN,
the success rate of verification, and average verification time of all the solved tasks.
Overall, the results show that \tool scales to larger QNNs with the attack radius $r$ up to 30.

Interestingly, we found that the verification cost is non-monotonic w.r.t. the attack radius.
To understand the reason, we analyze the distribution of (non-)robust samples that are solved by \tool.
The results on the QNNs P2-$y$ for $y\in\{4,6,8,10\}$ for each attack radius $r$ are shown in Figure~\ref{fig:dis_samples}.
The blue (resp.\ yellow) bars show the proportion (in percentage) of robust (resp.\ non-robust) samples among all the solved tasks.
The red dotted line reflects the change in the verification success rate.
From the upper two sub-figures in Figure~\ref{fig:dis_samples},  we can observe
that the success rate falls when the attack radius closing the robustness boundary and rises afterwards, conforming to the non-monotonicity in verification cost in Table~\ref{tab:scal_num}.
This indicates that the robustness of QNNs w.r.t. input regions with smaller or larger attack radii $r$ is relatively easier to prove/falsify, while it is harder for the attack radii close to the robustness boundary.
We also found that though the accuracy of these QNNs are similar (cf. Table~\ref{tab:bench}), their robustness are quite different, and particularly using more quantization bits results in an overall more robust QNN, i.e., more robust samples under the same attack radius. Similar results can be observed for P1-$y$ and P3-$y$ (cf. our website~\cite{website}).

\begin{tcolorbox}[left=0mm,right=0mm,top=0mm,bottom=0mm]
	\textbf{Answer to RQ2:}
	\tool is able to solve more verification tasks than the state-of-the-art verifier~\cite{smtQNN,scaleQNN21} within the same time and is two orders of magnitude faster on verification tasks that can be solved by~\cite{smtQNN,scaleQNN21}.
	\tool scales well to the QNNs with up to 894 neurons, attack radius 30 and quantization bit size from 4 to 10.
\end{tcolorbox}


\subsection{RQ3: \tool for Computing MRR}\label{subsec:MaxR_exp}
To answer {\bf RQ3}, we use \tool to compute the maximum robustness radii for QNNs P1-4 and P1-6.
We use the same 100 samples of MNIST as in Section~\ref{subsec:IA}, resulting in 200 computation tasks in total. Each task is run by \tool with 2 hours timeout.
Recall that it is essential to select appropriate values for {\tt StartR} and {\tt Step} parameters of Algorithm~\ref{alg:maxR}.
In this work, we conduct the experiments only to demonstrate the feasibility of \tool on MRR computation, hence we set the two parameters both as 10 directly.

\begin{figure}[t]
	\centering
	\includegraphics[width=.3\textwidth]{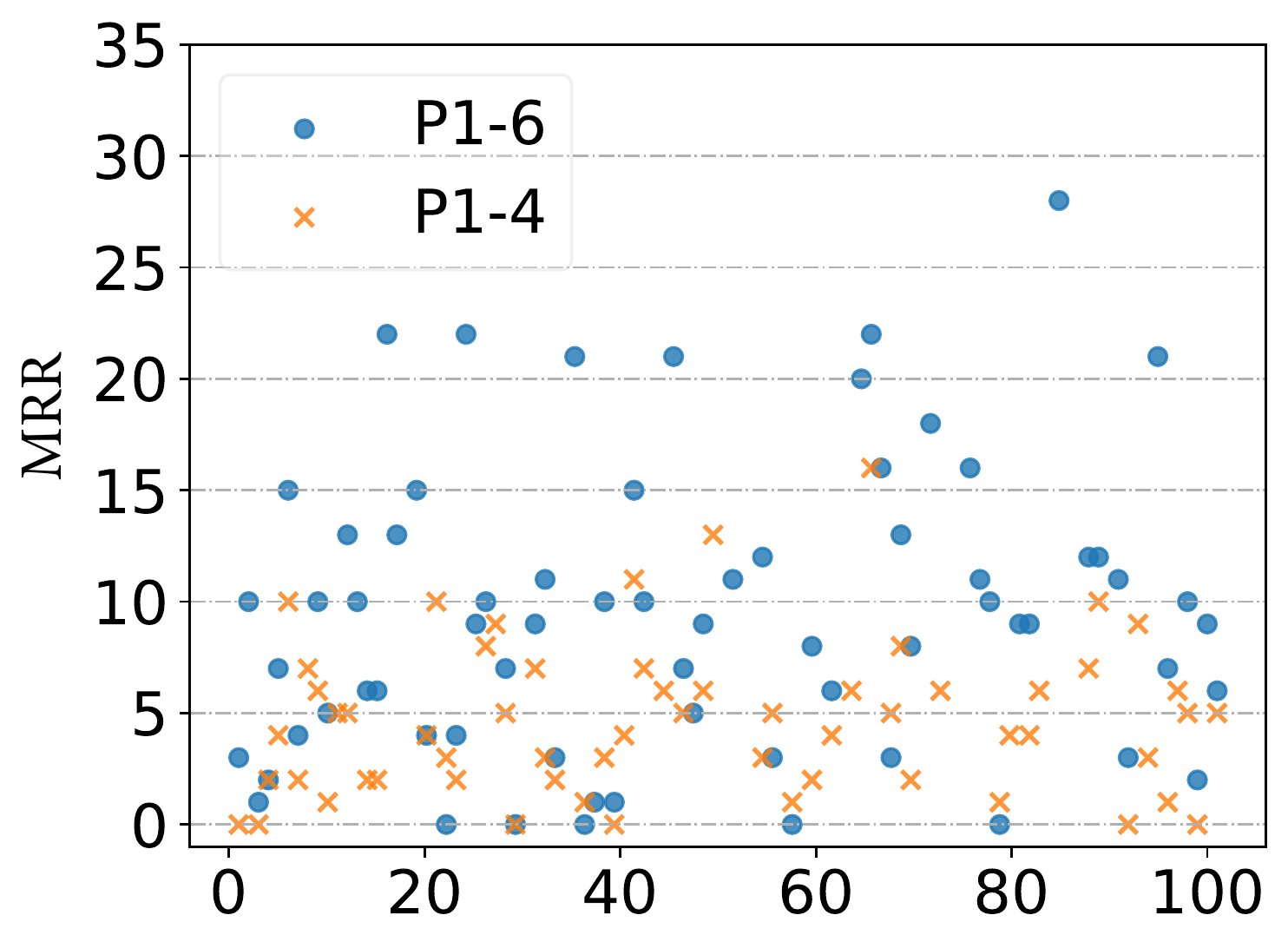}
	\vspace*{-3mm}
	\caption{Distribution of maximum robustness radii.\label{fig:MaxR}}
	\vspace*{-3mm}
\end{figure}

\tool successfully solved 68 of the 100 computation tasks for P1-6 with an average of 1,042 seconds per task,
and solved 59 of the 100 computation tasks for P1-4 with an average of 1,007 seconds per task.
Figure~\ref{fig:MaxR} shows the MRR distribution of  the solved tasks on two QNNs.
We can observe that the MRR distribution of P1-4 is more intensive with the average value 5
than the one of P1-6 with the average value 9.

The overall MRR of QNNs w.r.t. the same given set of samples also allow us to quantitatively measure and compare the robustness of QNNs. From the results, we can observe that P1-6 is more robust than P1-4 over the given set of 100 samples w.r.t. the distribution as well as the average MRR. We note that the comparison result would be more convincing by increasing the number of samples as well as their diversity.

\begin{tcolorbox}[left=0mm,right=0mm,top=0mm,bottom=0mm]
\textbf{Answer to RQ3:}
\tool is able to effectively and efficiently compute maximum robustness radii (i.e., MRR) for most samples, enabling
overall comparison of robustness of QNNs.
\end{tcolorbox}

\subsection{Threats to Validity}
Our approach is designed for verifying quantized neural networks typically
deployed in safety-critical resource-constrained devices, where the number of quantization bits and neurons
is not large. In this context, the completeness of the approach is significant.
However, large-scale neural networks in different application domains
may require abstraction techniques, at the cost of completeness.
In this work, we focus on quantized feed-forward ReLU networks, one of the most fundamental and common architectures.
It requires more experiments 
to confirm whether our approach still
outperforms the SMT-based method for other quantized neural networks.
Below, we discuss potential external and internal threats to the validity of our results. 

One of the potential external threats is the selection of evaluation subjects.
To mitigate this threat, we use 13 QNNs with the same or similar architectures as~\cite{GiacobbeHL20,scaleQNN21},
trained on two datasets, MNIST and Fashion-MNIST,
both of which are widely used for training QNNs/BNNs in the literature, e.g., \cite{narodytska2018verifying,NPAQ19,AWBK20,scaleQNN21,BDD4BNN,GiacobbeHL20}.
To be diverse, the QNNs involve 3 architectures and 4 quantization bit sizes, and the attack radius is set in a wide range (e.g., from 1 to 30 for QNNs trained on MNIST).

Another potential external threat is that the solving time of two ILP problems with similar sizes can be quite different.
Consequently, the verification time of the same QNN and same attack radius for different samples can be quite different.
To mitigate this threat, for each QNN and each attack radius, we evaluate our tool \tool on 50$\sim$100 samples,
showing that \tool performs well on most cases, and significantly outperforms the state-of-the-art verifier~\cite{smtQNN,scaleQNN21}.

The internal threat mainly comes from the interval analysis strategy used to reduce the size of constraints and verification cost while there is no theoretical guarantee. 
To mitigate this threat, we analyze the reduction of the number of Boolean variables and product terms in the encodings of the QNN P1-6 using 100 input samples with attack radius from 2 to 30,
resulting in 1400 encodings in total. The results show that interval analysis is very effective in most cases.
While the effectiveness of reducing the number of Boolean variables and product terms decreases with the increases of the attack radius, the verification cost can still be reduced significantly.

\section{Related Work}\label{sec:relatedwork}
There is a large and growing body of work on quality assurance techniques for (quantized) DNNs including testing (e.g.,~\cite{CW17b,MJZSXLCSLLZW18,MaLLZG18,PCYJ17,SunWRHKK18,TianPJR18,XieMJXCLZLYS19,compress2019,OdenaOAG19,XieMWLLL19,BCMS20,WebbRTK19,ZhangHML22,ZhaoCWYS021}) and formal verification (e.g., ~\cite{PT10,Ehl17,KBDJK17,KatzHIJLLSTWZDK19,GMDTCV18,HKWW17,SGPV19,WPWYJ18,TranLMYNXJ19,GokulanathanFMB20,AshokHKM20,ElboherGK20,YLLHWSXZ20,LiuSZW20,GuoWZZSW21,ZZCSCL22,LXSSXM22,}).
While testing techniques are often effective in finding
violations of properties, they cannot prove their absence;
formal verification and testing are normally complementary.
In this section, we mainly discuss the existing formal verification techniques for (quantized) DNNs,
which are classified into the following categories: constraint solving based, abstraction based, and decision diagram based.

\smallskip
\noindent{\bf Constraint solving based methods.}
Early work on formal verification of real numbered DNNs typically
reduces the problem to the solving of Satisfiability Modulo Theory (SMT) problem~\cite{KBDJK17,KatzHIJLLSTWZDK19,PT10} or
Mixed Integer Linear Programming (MILP) problem~\cite{ChengNR17,dutta2017output,fischetti2018deep},
that can be solved by off-the-shelf or dedicated solvers.
In theory, such techniques are both sound and complete unless abstraction techniques are adopted to improve scalability.
However, verification tools for real numbered DNNs cannot be used to guarantee the robustness of quantized DNNs (i.e., QNNs)
due to the fixed-point semantics of QNNs~\cite{GiacobbeHL20}.

Along this line, Narodytska et al.~\cite{narodytska2018verifying} proposed to reduce the verification problem of 1-bit quantized DNNs (i.e., BNNs) to
the satisfiability problem of Boolean formulas (SAT) or the solving of integer linear constraints.
Our QNN encoding  can be seen as a non-trivial generalization of their BNN encoding~\cite{narodytska2018verifying}.
Furthermore, we also propose to leverage interval analysis to effectively reduce the size of constraints and verification cost.
Using a similar encoding of \cite{narodytska2018verifying},
Baluta et al. proposed a PAC-style quantitative analysis framework for BNNs~\cite{NPAQ19} by leveraging approximate SAT model-counting solvers.
Jia and Rinard extended the encoding to three-valued BNNs~\cite{JiaR20}.
Narodytska et al. proposed a SAT-based verification-friendly BNN training framework~\cite{NarodytskaZGW20}.

Recently, accounting for the fixed-point semantics of QNNs,
Giacobbe et al.~\cite{GiacobbeHL20} pushed this direction further by introducing the first formal verification approach for multiple-bit quantized DNNs (i.e., QNNs).
They encode the verification problem of QNNs in first-order theory of quantifier-free
bit-vector with binary representation.
Later, first-order theory of fixed-point was proposed and used to verify QNNs~\cite{BaranowskiHLNR20}.
Henzinger et al.~\cite{scaleQNN21} explored several heuristics to improve efficiency and scalability of the SMT-based approach~\cite{GiacobbeHL20}.
To the best our knowledge, we are the first to reduce the verification problem of QNNs to the solving of integer linear constraints.
Experimental results show that our approach is significantly faster than the state-of-the-art~\cite{scaleQNN21}.

\smallskip
 \noindent{\bf Abstraction based methods.}
To improve efficiency and scalability, various abstraction techniques
have been used, which typically compute conservative bounds on the value ranges of the neurons for an input
region of interest. Abstract interpretation~\cite{CousotC77} has been widely used 
by exploring networks layer by layer~\cite{AndersonPDC19,GMDTCV18,LiLYCHZ19,LiLHY0ZXH20,SinghGPV19,SGMPV18,WPWYJ18,SGPV19,TranBXJ20,TranLMYNXJ19,YLLHWSXZ20}, typically with different abstract domains.
Almost all the existing work considered real numbered DNNs while few work  (e.g.,~\cite{SGPV19}) considered floating-point numbered DNNs.
The interval analysis
adopted in this work is an application of abstract interpretation with interval as its abstract domain.
It is an interesting future work to study other abstract interpretation based techniques 
to reduce the size of integer linear constraints. 
Another direction is to abstract neural networks~\cite{AshokHKM20,ElboherGK20,OBKatz22}, rendering them
suitable for formal verification. Abstraction based methods are sound but incomplete, so 
refinement techniques are normally needed to tighten bounds or refine
over-simplified networks.

While widely used for verifying DNNs, abstraction based verification of both QNNs and BNNs is very limited, except for the work \cite{scaleQNN21} which, similar to ours, is used to reduce the size of SMT formulas and verification cost.

Differential verification~\cite{LahiriMSH13} initially proposed for verifying a new version of a program w.r.t.\ a previous version,
has been applied to QNNs~\cite{PaulsenWW20,PaulsenWWW20,MohammadinejadP21}.
In general, they check if two (quantified) DNNs with the same architecture but different parameters output similar results (e.g., bounded by a small value) for each input from an input region.
As mentioned previously, these techniques cannot be used to directly and precisely verify the robustness of QNNs.

\smallskip
\noindent{\bf Decision diagram based methods.}
Decision diagram based verification methods were proposed for behavior analysis and verification of BNNs~\cite{CSSDvnn19,ddlearning19A,ddlearning19B,BDD4BNN}, e.g., allowing one to reason about the distribution of the adversarial examples or give an interpretation on the decision made by a BNN.
Choi et al.~\cite{CSSDvnn19} proposed a knowledge compilation based encoding that first transforms a BNN into a tractable Boolean circuit and then into a more tractable circuit, called Sentential Decision Diagram (SDD).
Based on the SDD representation, polynomial-time queries to SDD can be utilized to explain and verify the behaviors of BNNs efficiently.
In parallel, Shih et al. proposed a quantitative verification framework for BNNs~\cite{ddlearning19A,ddlearning19B},
where a given BNN with an input region of interest is modeled using a Binary Decision Diagram (BDD) by leveraging BDD-learning~\cite{Nakamura05}, which has limited scalability.
Very recently, Zhang et al.~\cite{BDD4BNN} proposed a novel BDD-based verification framework for BNNs, which exploits the internal structure of the BNNs to construct BDD models instead of BDD-learning. Specifically, they translated the input-output relation of blocks in BNNs to cardinality constraints which can then be encoded by BDDs.
Though decision diagram based methods enable behavior analysis and quantitative verification, they can hardly be extended to QNNs, i.e., multiple-bit quantized DNNs, due to the large space of QNNs.

\section{Conclusion}\label{sec:conc}
We have proposed the first ILP-based analysis approach for QNNs. We presented a prototype tool \tool and conducted thorough experiments on various QNNs with different quantization bit sizes. Experimental results showed that \tool is more efficient and scalable than the state-of-the-art, enabling the computation of maximum robustness radii for QNNs which can be used as a metric for robustness evaluation of QNNs.
We also found that the accuracy of QNNs stays similar under different quantization
bits, but the robustness can be greatly improved with more quantization bits, using quantization-aware training
while it is not using post-training quantization.

This work presented the first step towards the formal verification of QNNs. For future work, it would be interesting to investigate formal verification of QNNs that have more complicated architectures, activation functions, and a larger number of neurons.

\begin{acks}
This work is supported by the National Key Research Program (2020AAA0107800),
National Natural Science Foundation of China (NSFC) under Grants Nos. 62072309 and 61872340,
an oversea grant from the State Key Laboratory of Novel Software Technology, Nanjing University
(KFKT2022A03), Birkbeck BEI School Project (EFFECT), and the Ministry of Education, Singapore under its Academic Research Fund Tier 3 (Award ID: MOET32020-0004).
Any opinions, findings and conclusions or recommendations
expressed in this material are those of the authors and do not reflect the views of the Ministry of Education, Singapore.
\end{acks}

\balance


\end{document}